\documentclass{ieeeaccess}
\usepackage{cite}
\usepackage{amsmath,amssymb,amsfonts}
\usepackage{algorithmic}
\usepackage{graphicx}
\usepackage{textcomp}
\usepackage{url}
\usepackage{booktabs}
\usepackage{float} 

\usepackage{multirow}
\def\BibTeX{{\rm B\kern-.05em{\sc i\kern-.025em b}\kern-.08em
    T\kern-.1667em\lower.7ex\hbox{E}\kern-.125emX}}
\begin{document}
\history{Date of publication xxxx 00, 0000, date of current version xxxx 00, 0000.}
\doi{10.1109/ACCESS.2023.0322000}

\title{DeepFace-Attention: Multimodal Face Biometrics for Attention Estimation with Application to e-Learning}
\author{\uppercase{R}oberto \uppercase{D}aza\authorrefmark{1},
\uppercase{L}uis \uppercase{F}. \uppercase{G}omez\authorrefmark{1}, \uppercase{J}ulian \uppercase{F}ierrez\authorrefmark{1}, \uppercase{A}ythami \uppercase{M}orales\authorrefmark{1}, \uppercase{R}uben \uppercase{T}olosana\authorrefmark{1} and \uppercase{J}avier \uppercase{O}rtega-\uppercase{G}arcia\authorrefmark{1}
}

\address[1]{Biometrics and Data Pattern Analytics Laboratory, Universidad Autonoma de Madrid, Campus de Cantoblanco, Madrid 28049, Spain}

\tfootnote{ This work was supported in part by project HumanCAIC under Grant TED2021-131787B-I00 MICINN; in part by project BBforTAI under Grant PID2021-127641OB-I00 MICINN/FEDER; in part by the BIO-PROCTORING (GNOSS Program, Agreement Ministerio de Defensa-UAM-FUAM dated 29-03-2022); in part by the Catedra ENIA UAM-VERIDAS en IA Responsable (NextGenerationEU PRTR) under Grant TSI-100927-2023-2; and in part by the Autonomous Community of Madrid. The work of Roberto Daza was supported by the FPI Fellowship from MINECO/FEDER. The work of Aythami Morales was supported by the Madrid Government (Comunidad de Madrid-Spain) under the Multiannual Agreement with Universidad Autónoma de Madrid in the line of Excellence for the University Teaching Staff in the context of the Regional Program of Research and Technological Innovation (V PRICIT).
}

\markboth
{Roberto Daza \headeretal: DeepFace-Attention: Multimodal Face Biometrics for Attention Estimation with Application to e-Learning}
{Roberto Daza \headeretal: DeepFace-Attention: Multimodal Face Biometrics for Attention Estimation with Application to e-Learning}

\corresp{Corresponding author: Roberto Daza (e-mail: roberto.daza@uam.es).}

\begin{abstract}

This work introduces an innovative method for estimating attention levels (cognitive load) using an ensemble of facial analysis techniques applied to webcam videos. Our method is particularly useful, among others, in e-learning applications, so we trained, evaluated, and compared our approach on the mEBAL2 database, a public multi-modal database acquired in an e-learning environment. mEBAL2 comprises data from 60 users who performed 8 different tasks. These tasks varied in difficulty, leading to changes in their cognitive loads. Our approach adapts state-of-the-art facial analysis technologies to quantify the users' cognitive load in the form of high or low attention. Several behavioral signals and physiological processes related to the cognitive load are used, such as eyeblink, heart rate, facial action units, and head pose, among others. Furthermore, we conduct a study to understand which individual features obtain better results, the most efficient combinations, explore local and global features, and how temporary time intervals affect attention level estimation, among other aspects.
We find that global facial features are more appropriate for multimodal systems using score-level fusion, particularly as the temporal window increases. On the other hand, local features are more suitable for fusion through neural network training with score-level fusion approaches. Our method outperforms existing state-of-the-art accuracies using the public mEBAL2 benchmark.

\end{abstract}


\titlepgskip=-21pt

\maketitle

\section{Introduction}
\label{sec:introduction}
\PARstart{A}{ttention} is defined as the ability to focus, specifically, to exert on a conscious cognitive effort regarding a specific task or stimulus at a given moment~\cite{tang2022using, kahneman1973attention}. Therefore, it is used as a measure of the exerted effort. The level of attention can vary from a state of high attention, where a person is highly concentrated and experiences high levels of cognitive load and mental effort, to low levels, where a person is distracted or uninterested.

Attention estimation has proven to be of great value in important areas such as driver fatigue detection~\cite{bergasa2006real, 2019_IBPRIA_HRdriver_JHO}, advertising and product design~\cite{wedel2008eye}, mental health disorders~\cite{lim2012brain}, lie detection~\cite{leal2008blinking, mann2002suspects}, human-computer interfaces~\cite{iqbal2004task}, education~\cite{hernandez2019edbb}, etc.

Attention estimation is particularly valuable in e-learning environments~\cite{daza2021alebk, daza2023matt} because it offers feedback on students' cognitive and emotional states during online sessions. This is significant as attention is defined as the cognitive effort exerted on a task \cite{tang2022using} and plays a pivotal role in ensuring accurate comprehension during learning. In e-learning environments, there are challenges compared to face-to-face education, with one of the most important being the lack of direct contact between the teacher and the student. This results in the teacher being unaware of the student's study difficulties, like high or low levels of attention. Video-based attention estimation technologies overcome this limitation \cite{peng2020predicting}, representing a valuable tool to enhance both face-to-face and online education. 

\begin{figure}[t]
    \centering
    \includegraphics[width=\columnwidth]{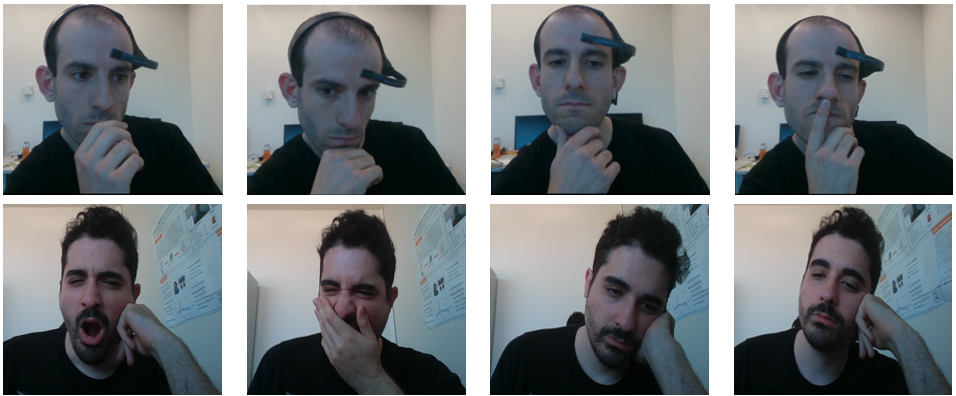} 
    \caption{Examples of different real students' attention levels during an e-learning session. (Top) High attention image sequence. (Bottom) Low attention image sequence.}
    \label{fig:Examples}
\end{figure}

Facial gestures often provide subtle indicators of an individual's attention level or cognitive load. When people are intensely focused or experiencing high cognitive demands, their facial expressions can change, reflecting the strain or concentration they are undergoing (see Fig.~\ref{fig:Examples}). Automatic attention estimation through image processing is a challenging task still under development. In this regard, the recent advances in face analysis techniques based on deep learning have also helped to improve attention estimation based on computer vision methods. The most advanced multimodal systems for attention estimation have reached around $80\%$ accuracy, outperforming the majority of existing monomodal systems~\cite{daza2023matt, zaletelj2017predicting}. Multimodal systems stand out for considering multiple variables that affect attention in the learning process, which allows a more global and complete perspective~\cite{2018_INFFUS_MCSreview1_Fierrez}.

Taking into consideration all of the above, the main contributions of the present paper are:
\begin{itemize}

\item We present a novel multimodal learning framework for attention estimation through image processing. This framework performs facial analysis to relate high and low levels of attention with behavior and physiological processes such as eyeblink, face gestures, and head pose, among others.

\item Our framework consists of 5 modules built on Convolutional Neural Networks (CNNs) that are trained to extract facial features that potentially correlate with attention. The most relevant modules for attention estimation and their effective combinations are identified within the e-learning context of the mEBAL2 database.

\item The results indicate that in multimodal attention estimation systems using score fusion, global features provide additional discriminating information compared to local features. However, multimodal attention estimation systems based on score fusion with neural network training generalize better with local features.

\item Our approach outperforms the state of the art, achieving a classification accuracy in attention level estimation of 85.92\%  on the mEBAL2 database.

\end{itemize}
A preliminary version of this article was presented in~\cite{daza2023matt}. This article significantly improves~\cite{daza2023matt} in various aspects:

\begin{itemize}

\item Compared to MATT~\cite{daza2023matt}, we now add a new module for heart rate estimation and study its relationship with attention estimation.

\item The mEBAL2~\cite{daza2024mebal2} database for attention estimation is used to train and evaluate the proposed system. In comparison with MATT~\cite{daza2023matt} (which used the first mEBAL version with $22$ users~\cite{daza2020mebal}), we now use the new version, mEBAL2, including $60$ students with approximately $1800$  minutes of video recordings. This represents a significant increase, with around 1140 additional minutes of recordings in comparison to ~\cite{daza2023matt}.

\item We add new comprehensive experiments including analysis of global and local features for each facial module.  We introduce a new method of score-level fusion through neural network training and a new architecture based on feature selection.

\item Unlike MATT~\cite{daza2023matt}, which utilized a one-minute time frame, we explored three time windows of $30$, $60$, and $120$ seconds.

\item Finally, our method outperforms the method presented in MATT~\cite{daza2023matt}, achieving an error reduction of $28.5\%$ in the mEBAL2 database.

\end{itemize}

The rest of the paper is organized as follows. Section $2$ summarizes works related to attention level estimation. Section $3$ describes the materials and methods, including the database, proposed technologies and features to estimate attention levels. Section $4$ presents the experiments and comparison with other state-of-the-art approaches. Finally, section $5$ provides conclusions and future investigations.

\section{Related Work}
\subsection{Brain activity measurement}

Attention estimation has been widely studied and currently there are different methods that come along with certain benefits and limitations~\cite{hall2020guyton}. Some of the most popular ones are:

\subsubsection{Electroencephalography (EEG)}

The EEG records the electrical activity of the brain through electrodes placed on the scalp. It measures neural activity by detecting changes in the voltage fluctuations generated by brain cells, specifically, the ones produced usually by synaptic excitations of the dendrites of pyramidal cells in the top layer of the brain cortex~\cite{kirschstein2009source, li2009towards}. The strength of the signals primarily relies on the synchronized firing of numerous neurons and fibers. Thousands or even millions of neurons are required to capture information effectively~\cite{hall2020guyton}. EEG data is recognized as one of the most efficient and unbiased approaches in estimating attention levels~\cite{chen2018effects, li2011real}, since these signals are sensitive to mental effort, cognitive demands, and mental states such as learning, deception, perception, and stress. Therefore, EEG provides real-time information about brain activity and it’s particularly useful for capturing quick changes in attention. EEG can be condensed to of five different signal types that reflect different mental states and activities. These signals are classified into different frequency bands: $\delta$ ($<4$Hz), $\theta$ ($4$-$8$ Hz), $\alpha$ ($8$-$13$ Hz), $\beta$ ($13$-$30$ Hz), and $\gamma$ ($>30$ Hz). However, the main disadvantage of this method is its intrusiveness, requiring precise tools to be placed on the student's head, which becomes impractical in e-learning environments with thousands of students.

\subsubsection{Physiological} 

This category associates attention with physiological responses like heart rate~\cite{haapalainen2010psycho, hernandez2020heart}, eyeblink~\cite{daza2021alebk,daza2023matt,daza2020mebal,daza2024mebal2}, eye pupil size ~\cite{rafiqi2015pupilware, krejtz2018eye}, electrodermal activity ~\cite{boucsein2012electrodermal}, etc. To measure these physiological signals and then correlate them with attention, specific sensors are used for each method that are then combined to obtain higher accuracy in attention estimation. 

\subsubsection{Behavior}

In comparison with the physiological category, this category analyzes the user’s noticeable patterns and behaviors to deduce attention levels. It’s based on external behavior observation that has proven to have a close relation with attention. Some of these behaviors are head pose~\cite{luo2022three, zaletelj2017predicting,raca2015translating, Becerra2024}, gaze tracking~\cite{wang2014eye, becerra2023m2lads}, facial expressions  ~\cite{monkaresi2016automated, mcdaniel2007facial, grafsgaard2013automatically}, physical actions that happen to be related with attention (e.g., leaning closer to the screen)~\cite{fujisawa2009estimation}, etc.

\subsection{Attention estimation methods based on image processing} \label{sec: State of art methods}

Here we employ images obtained from the webcam to infer the attention level of the users. The main advantage of this approach is that it doesn’t require specialized sensors more than a webcam, which makes it particularly attractive in areas like education, where accessibility is important. Currently there are monomodal systems like ALEBK~\cite{daza2021alebk} and multimodal ones like MATT~\cite{daza2023matt}. For example, MATT combines physiological and behavior estimations (pulse, facial analysis, etc). Multimodal systems have proven to be more efficient in attention estimation. 

The article~\cite{fridman2018cognitive} proposed $2$ monomodal methods to detect cognitive load in car driving environments. The used database defined $3$ states of cognitive load (high, medium, and low), which corresponded to variable difficulty activities (based on n-back task) that drivers had to perform; and the database had a total of $92$ users. The proposed methods were based on the eye state, starting with the first method that focused on the eye pupil’s position estimation (using face detection, landmark detection, etc) with Hidden Markov Models (HMMs) to estimate cognitive load. The second approach was based on Convolutional Neural Networks with $7$ convolutional layers, and the input was a temporally-stacked sequence of raw grayscale eye region images. The HMMs approach reached an average precision of $77.7\%$ while the CNN got $86.1\%$. The main issue was the cognitive load assumption without validating it using specific sensors, like EEG for example. 

ALEBk~\cite{daza2021alebk} represents a monomodal approach based on the relation between eyeblink and cognitive activity. Several studies have found clear evidence ~\cite{daza2020mebal,daza2021alebk, bagley1979effect, holland1972blinking} that lower eyeblink rates are associated with high attention levels, and vice versa. Based on this assumption, ALEBk~\cite{daza2021alebk} uses an eyeblink detector supported by convolutional neural networks to obtain the eyeblink frequency using RGB videos. With this information, the system classifies between high or low attention. The network was trained using the mEBAL~\cite{daza2020mebal} database with $22$ users performing tasks in an e-learning environment. Attention ground truth was obtained with an EEG band and the system reached a maximum accuracy ($1$-EER) of $70\%$ approximately.

The multimodal approach presented in~\cite{zaletelj2017predicting} used a Kinect One sensor to perform attention estimation. It only used behavior features, specifically gaze point, body posture and facial movements. The features were obtained from the signals of the Kinect SDK. This process included normalizing and filtering the signals using z-scores and an 11s-wide Gaussian filter. Subsequently, a 7-feature vector was selected by combining these signals. Finally, a $3$-level attention classification was made (low, medium, high) using different classifiers like decision tree, K-nearest neighbors, Subspace K-NN, etc. 
This study used a database captured in an e-learning environment of $18$ users with a length of $122$ minutes in total. 
The way how the  attention level ground truth was obtained is the main problem of this database, since it was through human observers, which can generate a lack of reliability in the results. Obtained results show a maximum accuracy of $75\%$ with a considerable variability between users. 

In~\cite{peng2020predicting}, a multimodal system is presented to estimate attention in a learning environment. This system extracted features from the face and also head movements, like mouth features (speaking or smiling), eye aspect ratio~\cite{soukupova2016eye}, leaning closer to the screen, etc, to estimate attention. It’s a simple system that uses a landmark detector to obtain the previously mentioned features from facial landmarks. Then, statistical measures like max, min, mean, variance, range and spectral entropy of face and head features are used for a random forest regression model, that predicts mean attention in a 10-second window. The used database consisted of recorded videos ($176$ minutes) of  $7$ middle school students while they interacted with an online tutoring system, along with EEG data. The authors reported an average RMSE of $12.66$ and indicated that both face and head movements provided useful information for attention estimation.

The authors of~\cite{goldberg2021attentive} proposed a multimodal attention estimation system for classrooms with several students to improve learning. The artificial vision approach used features like head pose, gaze direction and facial expression (facial action units) obtained with OpenFace~\cite{baltrusaitis2018openface} and regression models to estimate attention were trained with them. The approach classified the student’s commitment level as “attentive” and “non-attentive” in one-second time frames. The database was obtained from university seminars with $52$ students recorded with $3$ cameras, even though the automatic approach used only 30 users. The attention level labels used as ground truth were obtained by evaluators that observed each student’s behavior throughout sessions. The head pose feature got the least correlation regarding manual scores, and the highest correlation was reached with the combination of all $3$ modules (r=$0.61$).

\begin{table}
\centering
\setlength\tabcolsep{7pt}
 \renewcommand{\arraystretch}{1.1}
  \caption{mEBAL2 Database: Sensors.}
  \label{tab:mEBal_sensors}
  \renewcommand{\arraystretch}{1}

\begin{tabular}{lc}
\toprule
\textbf{Sensors} & \multicolumn{1}{l}{\textbf{Sampling Rate}} \\
\midrule
EEG Band\footnotemark[1] & $1$ Hz \\
1 RGB\footnotemark[2] camera & $30$ Hz \\
2 NIR\footnotemark[2] cameras & $30$ Hz \\
\bottomrule
\end{tabular}
\end{table}

\footnotetext[1]{\url{https://store.neurosky.com/pages/mindwave}}
\footnotetext[2]{\url{https://www.intelrealsense.com/wp-content/uploads/2020/06/Intel-RealSense-D400-Series-Datasheet-June-2020.pdf}}

MATT~\cite{daza2023matt} represents a multimodal approach  that uses a simple webcam and it’s based on different Convolutional Neural Network modules that extract behavior and physiological  features (head pose, eyeblink, facial action units, etc). For each module, a Support Vector Machine (SVM) is used as a binary classifier to determine high or low attention levels and at the end, all modules are combined with a score sum. Similar to ALEBk~\cite{daza2021alebk}, this approach was trained and evaluated on mEBAL~\cite{daza2020mebal} database with $22$ users and obtained a maximum accuracy ($1$-EER) of $82\%$ approximately. 

\subsection{Multimodal Machine Learning}


Multimodal systems have demonstrated great potential to improve the performance of unimodal systems \cite{2018_INFFUS_MCSreview1_Fierrez, fierrez2018multiple}, due to their enhanced comprehension capabilities. By integrating various data sources, these systems can leverage the redundancy and complementarity of information to achieve more accurate and robust results. Specifically, in attention estimation, analyzing a single facial feature category is typically not discriminative enough to classify attention levels \cite{daza2021alebk, daza2023matt, zaletelj2017predicting, peng2020predicting}. In contrast, multimodal systems show superior performance in estimating attention by integrating different unimodal systems based on diverse facial categories, such as eyeblinks and heart rate \cite{daza2023matt, zaletelj2017predicting, peng2020predicting}. Various fusion strategies have been proposed in the literature \cite{2018_INFFUS_MCSreview1_Fierrez, fierrez2018multiple, leng2017dual, LENG20131, hong2023decoupled, yao2023extended, daza2024mebal2}, including feature level fusion, score level fusion, and model level fusion.
Feature level fusion involves combining data or signals at the feature level before they are input into a classification or regression model \cite{leng2017dual,atrey2010multimodal,2018_INFFUS_MCSreview1_Fierrez, fierrez2018multiple}. Leng et al. \cite{leng2017dual} employed Dual-Source Discriminative Power Analysis (DDPA) to assess the discriminative power of features from two different information sources, based on inter-class and intra-class variation, and subsequently fused them. Score level fusion, on the other hand, involves combining outputs from multiple models to reach a final decision. Various strategies are employed, including score sum, weighted sum, and voting, among others \cite{2018_INFFUS_MCSreview1_Fierrez, atrey2010multimodal, LENG20131}. Other works \cite{yao2023extended, hong2023decoupled, daza2024mebal2} have implemented model level fusion. Yao et al. \cite{yao2023extended} proposed an extension of the conventional Vision Transformer (ViT). This approach applied a strategy for fusing through a structure that integrates extended visual transformers and Cross-Modality Attention (CMA), thus incorporating modality fusion directly into the model processing stages.

\section{MATERIALS AND METHODS}
\label{sec:MATERIALS AND METHODS}

\subsection{Database}

To carry out this study, we selected the public database mEBAL2~\cite{daza2024mebal2}, a Multimodal Database for EyeBlink Detection and Attention Level Estimation. It’s the first database that we’re aware of being captured in an e-learning environment, providing information on attention levels and eyeblink samples. mEBAL2 is a public database obtained in a real e-learning environment using the research platform edBB~\cite{hernandez2019edbb,daza2023edbb, becerra2023m2lads}. We used this database, which includes data from $60$ students who performed various carefully designed tasks to induce changes in cognitive load. These tasks were designed to induce changes in students' attention and evaluate the cognitive load associated with each situation. Among the tasks included in the acquisition protocol, the task of committing fraud/copying was included, as previous research demonstrated that this activity requires a higher cognitive load~\cite{mann2002suspects}. Students were presented with diverse scenarios to engage in copying responses, like using different electronic devices (mobile phones, laptops), employing "cheat sheets," interacting with peers to obtain answers, and more. The database also induced an altered state in the students, to observe how it affected their attention during the e-learning session. During a specific moment, students engage in physical exercise, inducing an altered state that affects their heart rate, simulating a state of nervousness/stress. Afterward, they resume the session.

mEBAL2 contains signals from multiple sensors, including face video and electroencephalogram (EEG) data. The data was captured with the following sensors (see Table~\ref{tab:mEBal_sensors}): An Intel RealSense composed of 1 RGB camera and 2 NIR cameras, along with an EEG band provided by NeuroSky. It is worth mentioning that previous studies have also utilized this EEG headset to gather EEG and attention signals~\cite{rebolledo2009assessing,li2009towards,lin2018mental}, as EEG  measurement is considered one of the most effective methods for attention estimation. The information from the EEG band includes $5$ EEG signals ($\delta$, $\theta$, $\alpha$, $\beta$, $\gamma$). Through the official NeuroSky SDK, mEBAL2 includes information regarding attention and meditation level, and a temporal sequence with eyeblink strength. Attention and meditation levels are assigned values ranging from $0$ to $100$. We employed the attention levels acquired from the EEG headset as ground truth to both train and evaluate our image-based attention level estimation approach.
Additionally, mEBAL2~\cite{daza2024mebal2} provides $10550$ eyeblink samples, the largest existing public eyeblink database for research.


\begin{figure}[t]
    \centering
    \includegraphics[width=\columnwidth]{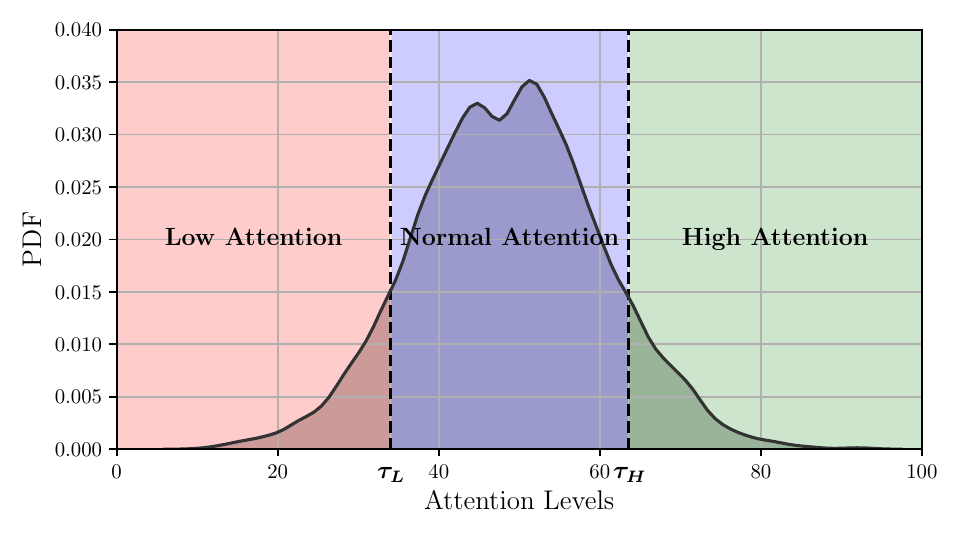} 
    \caption{Probability Density Function of obtained attention with EEG band from 60 students in the mEBAL2 database~\cite{daza2024mebal2}, along with our attention levels classification (high, normal, low) with used thresholds ($\tau_L$, $\tau_H$).}
    \label{fig:Dist_Y}
\end{figure}

To summarize, mEBAL2~\cite{daza2024mebal2} includes data from $60$ students who participated in e-learning sessions that lasted between $15$ to $30$ minutes. These sessions consisted of various activities related to mental load, visual attention, etc., such as filling in registration forms, answering logical and multiple-choice questions, performing visual exercises (describing images, finding differences), and more. Additionally, some of the students took part in events related to changes in attention, such as fraud/copying, physical exercise (see \cite{daza2023edbb} for a video demonstration\footnote[3]{\url{https://www.youtube.com/watch?v=JbcL2N4YcDM}}). All participants gave written informed consent. The study is in accordance with the Declaration of Helsinki.

Fig. \ref{fig:Dist_Y} shows the Probability Density Function of the attention levels of the $60$ students, with an average attention level around $50\%$, and the most frequent attention level being $55\%$.

\subsection{Face Analysis Modules} \label{sec:FaceAnalysisModule}

\begin{figure*}[t!]
\centering
\includegraphics[width=\textwidth]{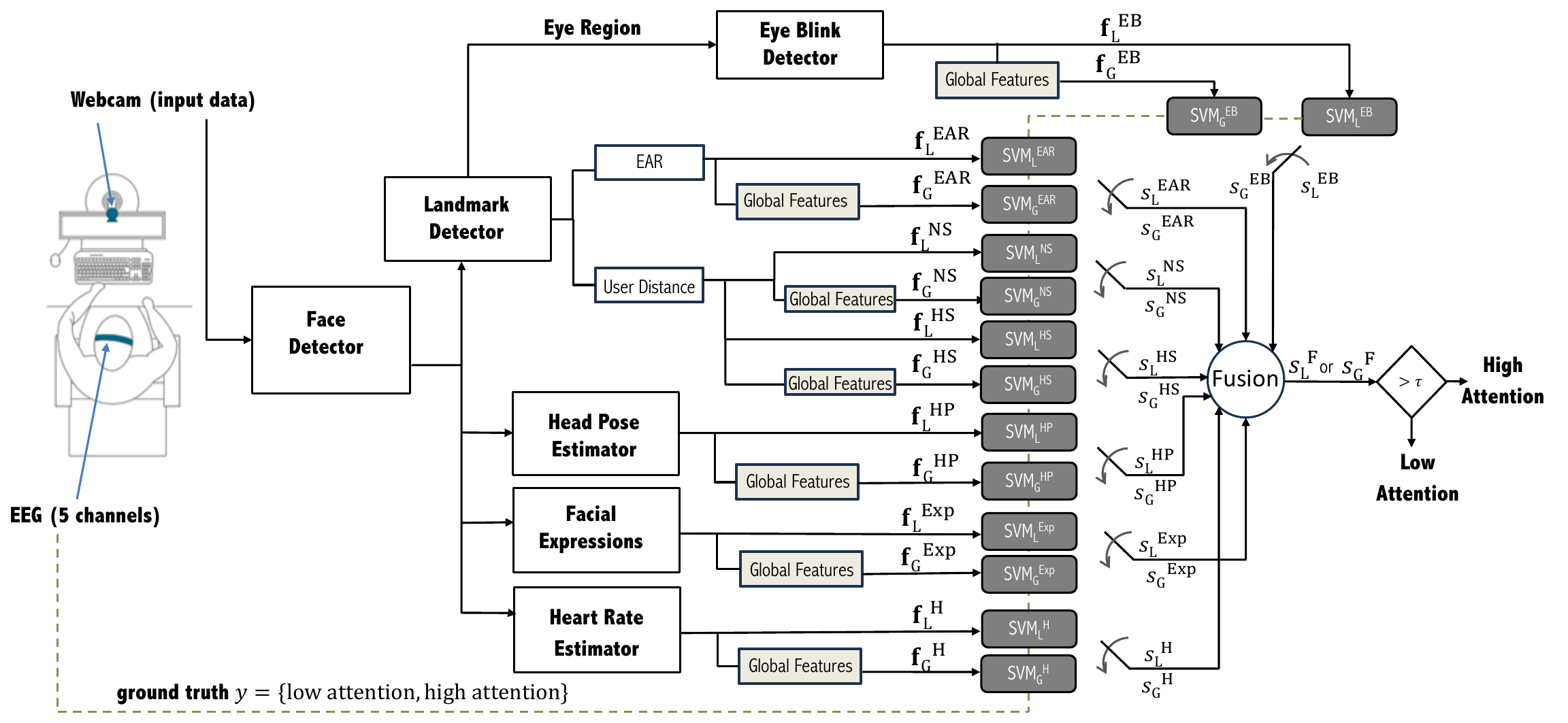} 
\caption{Block diagram of the proposed multimodal approach for attention estimation (DeepFace-Attention). The dashed line represents the ground truth used for training the SVMs. The two strategies used, global features ($\textbf{f}_{\textnormal{G}}$) and local features ($\textbf{f}_{\textnormal{L}}$), are shown. The feature vectors from each module are denoted as $\mathbf{f}^{\mathnormal{y}}_{\mathnormal{x}}$,  and the score for each SVM is denoted as ${\mathnormal{s}_{\mathnormal{x}}^\mathnormal{y}}$. Here, ${\mathnormal{x} \in \{\textnormal{L}, \textnormal{G}\}}$ specifies whether the features are global or local, and $\mathnormal{y}$ represents the facial feature category, $\mathnormal{y} \in \{\textnormal{EB}, \textnormal{HP}, \textnormal{EAR}, \ldots\}$. Finally, $\mathnormal{s}^{\textnormal{F}}$ represents the fusion of scores.}
\label{Block_diagram}
\end{figure*}

Our proposed DeepFace-Attention estimates attention through the facial analysis of images captured by a webcam. Different modules based on convolutional networks are used to extract facial features based on behavior as well as physiological signals, which have proven to estimate attention \cite{daza2023matt, daza2024mebal2, zaletelj2017predicting, yao2022semi}. Fig.~\ref{Block_diagram} shows our proposed system of attention estimation. The used modules are as follows: 

\textbf{Face Detection Module:} Our approach detects $2$D facial images using a state-of-the-art RetinaFace Detector~\cite{deng2020retinaface}. This robust single-stage face detector was trained using the Wider Face dataset~\cite{yang2016wider}. Once the facial position in the image is obtained, it is used as input for the subsequent modules.

\textbf{Landmark Detection Module:} We use the SAN landmark detector~\cite{dong2018style} to acquire facial landmarks, which comprises a $68$-landmark detection system based on VGG-$16$ plus $2$ convolutional layers trained on the $300$-W dataset~\cite{sagonas2016300}. The facial landmarks serve as a dual purpose in our approach. Firstly, these landmarks are used to extract facial features that have demonstrated relevance in attention estimation. Secondly, they are employed to locate the eye region of interest, which subsequently serves as input to the EyeBlink module.

Through facial landmarks, we obtain features related to attention estimation. Firstly, we focus on the eye state, specifically the Eye Aspect Ratio (EAR)~\cite{soukupova2016eye}  for each eye, which is related to the eye opening.

\begin{figure}[H]
    \centering
    \includegraphics[width=\columnwidth]{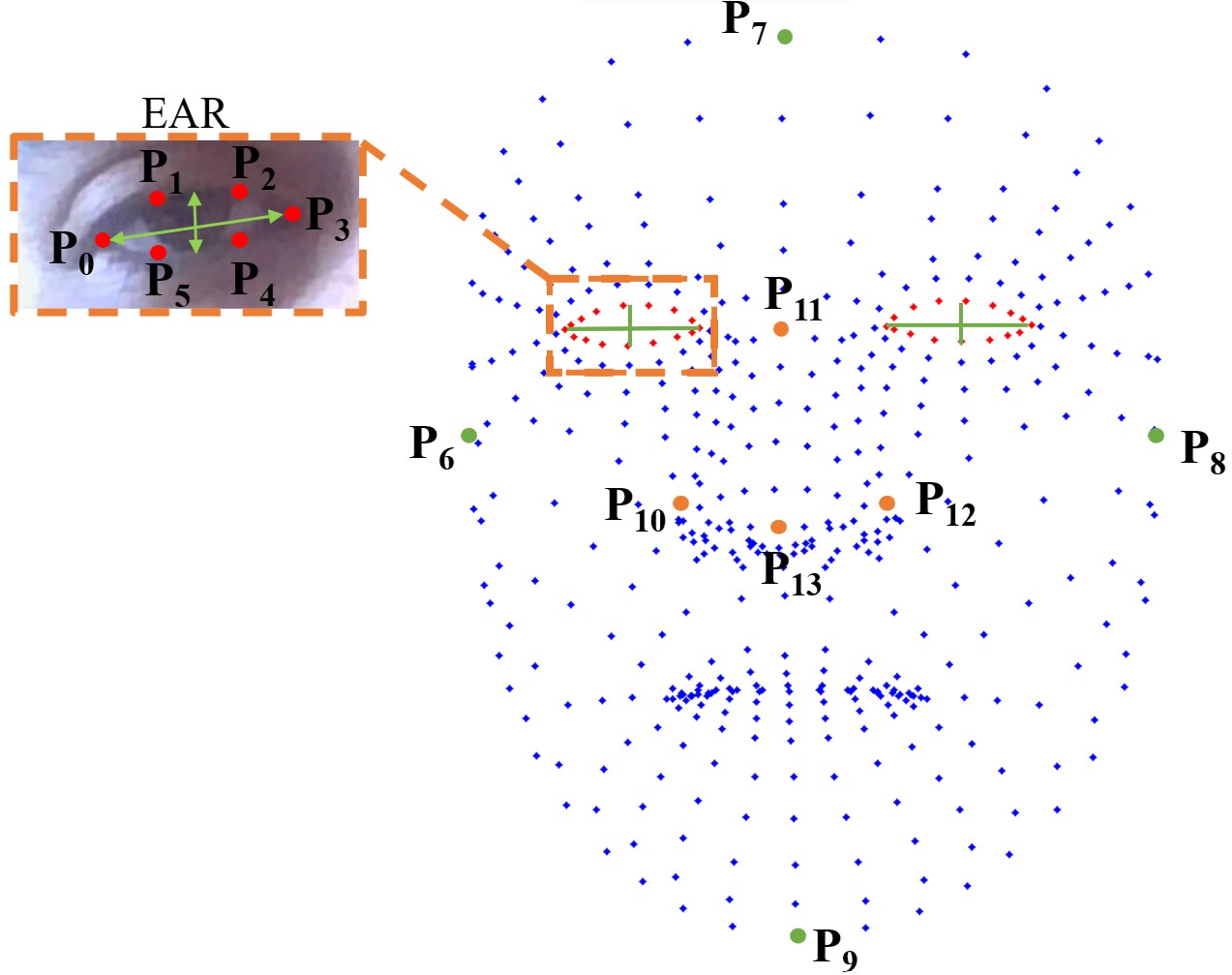} 
    \caption{Feature extraction from the Landmark Detection module. On the right eye, we show Eye Aspect Ratio (EAR) calculations. We also display the landmarks used to extract the width and height of the nose and head.}
    \label{Landamarks}
\end{figure}

The EAR is calculated following the next equation:


\begin{equation}
    \textrm{EAR} = \frac{\left \| \mathbf{P}_1- \mathbf{P}_5 \right\| + \left \| \mathbf{P}_2- \mathbf{P}_4 \right\|}{2 \left \| \mathbf{P}_0- \mathbf{P}_3 \right\|}
\end{equation}

\noindent where $\mathbf{P}_0$, $\ldots$, $\mathbf{P}_5$
 are the eye landmarks shown in Fig.~\ref{Landamarks}. The denominator is multiplied by $2$ because only one distance is calculated for horizontal eye landmarks.

We calculate the EAR parameter for each eye, so, two EAR features are obtained per frame.


The other 4 features are related to the student's distance from the screen, as previous studies have shown its usefulness in attention estimation~\cite{peng2020predicting}. We obtain the Width and Height of the Head and the Nose by simply subtracting the following landmarks: 
\begin{equation}
\textrm{H}_{\mathrm{W}} = P_{8_{\mathrm{x}}} - P_{6_{\mathrm{x}}}
\end{equation}
\begin{equation}
\textrm{H}_{\mathrm{H}} = P_{9_{\mathrm{y}}} - P_{7_{\mathrm{y}}}
\end{equation}
\begin{equation}
\textrm{N}_{\mathrm{W}} = P_{12_{\mathrm{x}}} - P_{10_{\mathrm{x}}}
\end{equation}
\begin{equation}
\textrm{N}_{\mathrm{H}} = P_{13_{\mathrm{y}}} - P_{11_{\mathrm{y}}}
\end{equation}

where $\mathbf{P}_{6}, \ldots, \mathbf{P}_{\mathrm{13}}$ are the eye landmarks shown in Fig.~\ref{Landamarks}.

Finally, we normalize all the values using z-score \cite{Fierrez-Aguilar2005_ScoreNormalization}, resulting in four features for each frame corresponding to the facial feature categories of Head Size (HS) and Nose Size (NS).

This landmark processing is in line with our previous works, see \cite{2023_PLOS_FacialParkinson_Gomez, 2021_CVPRW_Parkinson_Gomez} for more details.

\textbf{Head Pose Estimation Module:} The head pose is estimated using $2$D facial images obtained from the facial detection module. To achieve a balance between speed and precision, we used a Convolutional Neural Network (ConvNet) based on ~\cite{berral2021realheponet}. This head pose estimator was trained with data from the Pointing $04$~\cite{gourier2004estimating}  and Annotated Facial Landmarks in the wild~\cite{koestinger2011annotated} databases. This architecture calculates the vertical (pitch) and horizontal (yaw) angles, enabling us to infer the $3$D head pose from $2$D facial images. This module obtains the two angles that define the $3$D head pose for each frame, forming the facial feature category Head Pose (HP).


\textbf{EyeBlink Detection Module:}  The eye state has proven to be one of the most relevant indicators for attention estimation. We use an eye state classifier on each RGB frame, distinguishing between "open" or "closed" states, which is commonly employed as a blink detector in frame sequences. Our architecture is based on the approach presented in ALEBk~\cite{daza2021alebk}, and we trained it from scratch using the mEBAL database~\cite{daza2020mebal}, with RGB images only. The output values range between $0$ and $1$ and the input consists of two cropped images of the right and left eye. We apply the following approach to obtain the region of interest: \textit{i)} face detection, \textit{ii)} landmark detection, \textit{iii)} face alignment using the Dlib library, \textit{iv)} data quality assessment: we use the detectors' probabilities to evaluate the ROI quality from which we decide to maintain or not the alignment, or discarding the frame, and \textit{v)} eye cropping: we crop the region of each eye and resize it to $50\times50$.
This module obtains a value between $0$ and $1$ as a feature per frame for the facial feature category  EyeBlink (EB).
\begin{table*}[t]
\centering
\setlength\tabcolsep{7pt}
\renewcommand{\arraystretch}{1.5}
\caption{Features extracted from the face analysis modules for our proposed system. $W_l$ is the time window size (in seconds) analyzed to extract global or local features.}
\label{tab:Feature_Table}

\begin{tabular}{l l l l l}
\toprule
\textbf{Modules} & \textbf{Feature Categories}  & \textbf{Local Feature Vectors} & \textbf{Global Feature Vectors} \\
\midrule
Landmark    & \begin{tabular}[c]{@{}l@{}}EAR\\ HS\\ NS\end{tabular} & \begin{tabular}[c]{@{}l@{}}$\mathbf{f}^{\mathrm{EAR}}_{\mathrm{L}} \in \mathbb{R}^{2 \times W_{l}}$\\ $\mathbf{f}^{\mathrm{HS}}_{\mathrm{L}} \in \mathbb{R}^{2 \times W_{l}}$\\ $\mathbf{f}^{\mathrm{NS}}_{\mathrm{L}} \in \mathbb{R}^{2 \times W_{l}}$\end{tabular} & \begin{tabular}[c]{@{}l@{}}$\mathbf{f}^{\mathrm{EAR}}_{\mathrm{G}} \in \mathbb{R}^{2 \times 28}$\\ $\mathbf{f}^{\mathrm{HS}}_{\mathrm{G}} \in \mathbb{R}^{2 \times 28}$\\ $\mathbf{f}^{\mathrm{NS}}_{\mathrm{G}} \in \mathbb{R}^{2 \times 28}$\end{tabular} \\

Head Pose       & HP                      & $\mathbf{f}^{\mathrm{HP}}_{\mathrm{L}} \in \mathbb{R}^{2 \times W_{l}}$           & $\mathbf{f}^{\mathrm{HP}}_{\mathrm{G}} \in \mathbb{R}^{2 \times 28}$            \\
EyeBlink        & EB                    & $\mathbf{f}^{\mathrm{EB}}_{\mathrm{L}} \in \mathbb{R}^{1 \times W_{l}}$        & $\mathbf{f}^{\mathrm{EB}}_{\mathrm{G}} \in \mathbb{R}^{1 \times 28}$            \\
Facial Expression      & Exp                    & $\mathbf{f}^{\mathrm{Exp}}_{\mathrm{L}}  \in \mathbb{R}^{16 \times W_{l}}$           & $\mathbf{f}^{\mathrm{Exp}}_{\mathrm{G}}  \in \mathbb{R}^{16 \times 28}$            \\
Heart Rate   & H                          & $\mathbf{f}^{\mathrm{H}}_{\mathrm{L}} \in \mathbb{R}^{1 \times W_{l}}$       & $\mathbf{f}^{\mathrm{H}}_{\mathrm{G}} \in \mathbb{R}^{1 \times 28}$       \\
\bottomrule
\end{tabular}
\end{table*}

\begin{table*}[ht!]
    \centering
    \caption{Description of the $g_n^{k}$ ($k=1,\ldots,28$) global features of the global vector $\mathrm{g}_n$ extracted from each time series used in this work. Adapted from ~\cite{fierrez2005line, tolosana2015feature}.}
    \label{tab:GlobalFeatures}
        \begin{tabular}{lllll}
            \noalign{\hrule height 1pt}
            \textbf{\#} & \textbf{Feature Description}                                              & \textbf{\#} & \textbf{Feature Description}                &  \\
            \noalign{\hrule height 1pt}
            $g_n^{1}$  & $\text{Total positive velocity } \sum_{\left (\textbf{v}> 0 \right )}$       & $g_n^{2}$  & $\text{Total negative velocity } \sum_{\left (\textbf{v} < 0 \right )}$ &  \\
            $g_n^{3}$  & $(\text{1st maximum location in } \textbf{x})$                               & $g_n^{4}$  & $(\text{2nd maximum location in } \textbf{x})$        &  \\
            $g_n^{5}$  & $(\text{3rd maximum location in } \textbf{x})$                               & $g_n^{6}$  & $(\text{average velocity }  \tilde{\textbf{v}}) / \left | \textbf{v} \right |_{\max}$ &  \\
            $g_n^{7}$  & $(\text{average velocity } \tilde{\textbf{v}}) / \textbf{v}_{\max}$          & $g_n^{8}$  & $(\text{RMS velocity } \textbf{v}_{\text{RMS}}) / \left | \textbf{v} \right |_{\max}$      &  \\
            $g_n^{9}$  & $(\text{RMS centripetal acceleration }\textbf{a}_{c_{\text{RMS}}})/\left|\textbf{a}\right|_{\max}$   & $g_n^{10}$  &$(\text{RMS tangential acceleration } \textbf{a}_{t_{\text{RMS}}}) / \left |\textbf{a} \right |_{\max}$                                         &  \\
            $g_n^{11}$  & $(\text{RMS acceleration } \textbf{a}_{\text{RMS}}) / \left |\textbf{a}\right |_{\max} $               & $g_n^{12}$  & $(\text{average abs. centripetal acceleration }  \tilde{\left |\textbf{a}_{c}\right |}) / \left |\textbf{a}\right |_{\max}$                                         &  \\
            $g_n^{13}$  & $\text{standard deviation of velocity } \sigma_{\textbf{v}}$                & $g_n^{14}$  & $\text{standard deviation of acceleration } \sigma_{\textbf{a}}$ &  \\
            $g_n^{15}$  & $\text{average abs. jerk }  \tilde{\left | \textbf{j}  \right |} $          & $g_n^{16}$  & $\text{average jerk }  \tilde{\textbf{j}}$        &  \\
            $g_n^{17}$  & $ \text{maximum abs. jerk }\left | \textbf{j}  \right | _{\max}$                            & $g_n^{18}$  & $ \text{maximum jerk } \textbf{j}_{\max}$ &  \\
            $g_n^{19}$  & $\text{RMS jerk } \textbf{j}_{\text{RMS}}$                          & $g_n^{20}$  & $(\text{time of } \left | \textbf{j}  \right | _{\max})$  &  \\
            $g_n^{21}$  & $(\text{time of } \tilde{\textbf{j}}_{\max})$                               & $g_n^{22}$  & $\text{Total sign changes of } \textbf{v}$        &  \\
            $g_n^{23}$  & $(\sum_{(\textbf{v}>0)}{\left | \textbf{v} \right |})/(\sum_{(\textbf{v}<0)}\left | \textbf{v} \right |)$                       & $g_n^{24}$  & $(\sum_{(\textbf{v}>0)})/(\sum_{(\textbf{v}<0)})$              &  \\
            $g_n^{25}$  & $\textbf{x}_{\max}-\textbf{x}_{\min}$         & $g_n^{26}$  & $ \tilde{\textbf{v}} / (\textbf{x}_{\max}-\textbf{x}_{\min}) $      &  \\
            $g_n^{27}$  & $(\text{Total of local maximum in } \textbf{x})$           & $g_n^{28}$  & $(\text{average acceleration } \tilde{\left |\textbf{a}\right |}) $                                         & \\
            \noalign{\hrule height 1pt}
        \end{tabular}%
\end{table*}

\textbf{Facial Expression Module:} This module is based on the work by Zhang et al.~\cite{zhang2021learning}, who created a new architecture based on the subtraction of two embeddings to extract a disentangled feature space where the facial expression embedding was compacted, and the user's identity was ignored. The two branches are two FaceNet-Inception architectures pretrained with VGGFace2, where the first branch is fixed to preserve the identity information and the second branch is retrained with Google Facial Expression Comparison (FEC) dataset~\cite{vemulapalli2019compact} to improve the facial expression features. The model follows the same experimental protocol proposed in~\cite{zhang2021learning} using the triplet loss function to obtain the disentangled facial expression space. The result is 16 features per frame for the facial feature category of Facial Expression (Exp).



\textbf{Heart Rate Detection Module:} We employ the DeepPhys model to estimate the human heart rate using remote photoplethysmography (rPPG) based on the facial video sequences. This model is based on the  Convolutional Attention Network created by Chen and McDuff in~\cite{chen2018deepphys} and implemented by Hernandez-Ortega et al. in~\cite{hernandez2020comparative}, where the  DeepPhys architecture was trained on the COHFACE database~\cite{heusch2017reproducible}. The model comprises two parallel Convolutional Neuronal Networks branches that extract temporal and spatial information from videos: (i) Motion branch designed to realize a short-time video analysis to detect pixel changes over the scene, and (ii) Appearance branch designed to create attention masks based on the subject's appearance to help the motion model. This module outputs $\textbf{f}^{\mathrm{H}} \in \mathbb{R}^{1 \times W_{l}}$, which corresponds to a heart-rate estimation every second of the time window at hand (of size $W_{l}$ seconds).

%

\subsection{Feature Extraction Approaches: Local vs Global} \label{sec:Features}

Considering that the analysis of long temporal sequences increases the complexity of classification algorithms based directly on the time sequences, here we study to what extent are useful and efficient global features that integrate the information across time. To integrate the temporal information from the video sequences, we have adapted the global features proposed in~\cite{fierrez2005line,  2022_INFFUS_Indicators_Tato}.

The face analysis modules presented in the previous section are used to extract local and global features (see Table~\ref{tab:Feature_Table}). We then apply two different feature processing approaches for the extraction of local and global relationships.

First, to characterize the local relations we use the method presented in MATT~\cite{daza2023matt}. The features obtained from each module, denoted as $\textrm{\textbf{f}}_{x,y}$, where ${x \in \{1,\ldots,N\}}$ represents the specific feature and  $y$ represents the facial feature category {$y \in \{\mathrm{EB}, \mathrm{HP}, \mathrm{EAR}, \ldots\}$}, are used to obtain local feature vectors. For each facial feature category, a local feature vector is generated, capturing the changes in the facial attributes for high and low attention, as follows: \textit{i)} the facial analysis module’s features $\textrm{\textbf{f}}_{x,y}$ are averaged for each second of video, generating 
$\bar{\textrm{\textbf{f}}}_{x,y}$, and \textit{ii)} for each facial feature category, a local feature vector  $\textbf{f}^{y}_{\mathrm{L}} \in \mathbb{R}^{N \times W_{l}}$ is obtained by concatenating the 1s averages $\bar{\textbf{f}}_{x,y}$ across the time window of size  $W_l$ (30, 60, or 120 seconds), making $N \times W_l$ the dimension of the vector, where $N$ is the number of features per second. These local feature vectors are used to estimate the attention level every second.


Second, the characterization of global relationships proposed in this work (one of the novelties here in DeepFace-Attention with respect to MATT \cite{daza2023matt}) involves extracting statistical features from the outputs of the face analysis modules, which have previously demonstrated their effectiveness in other classification tasks~\cite{fierrez2005line, 2022_INFFUS_Indicators_Tato}. For each facial feature category, a global feature vector $\textrm{\textbf{f}}^{y}_{\mathrm{G}}$
is extracted from a sequence of features $\textbf{f}^{y}_{\mathrm{L}} \in \mathbb{R}^{N \times W_{l}}$, where $W_l$ is the time window size (in seconds) and  $N$ is the number of features per second. This sequence $\textbf{f}^{y}_{\mathrm{L}}$ is formed as before in the local representation by concatenating the 1s averages $\bar{\textbf{f}}_{x,y}$. The global feature vector $\textrm{\textbf{f}}^{y}_{\mathrm{G}}$ for each feature category $y \in \{\mathrm{EB}, \mathrm{HP}, \mathrm{EAR}, \ldots\}$ is now defined as a set $\textbf{g}_n\in \mathbb{R}^{28}$ with {$n \in \{1,2,\ldots,N\}$} where $N$ is the number of features per second as described in Table~\ref{tab:GlobalFeatures}.

\begin{figure*}[t!]
\centering
\includegraphics[width=\textwidth]{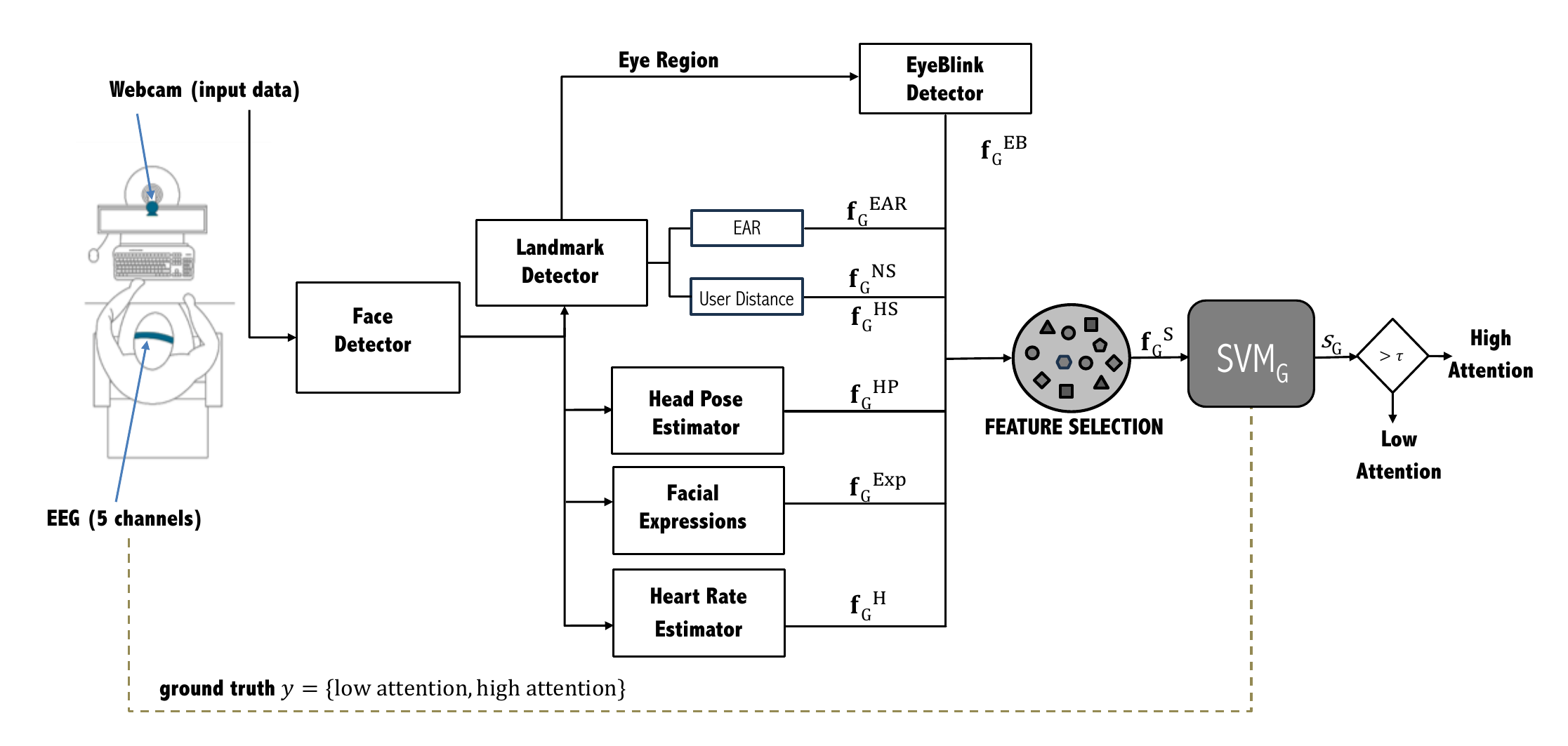} 
\caption{Block diagram using an approach of selection and fusion of global features for attention estimation. The dashed line represents the ground truth used for training the SVM. The global feature vector is denoted as $\mathbf{f}^{\mathnormal{y}}_{\textnormal{G}}$, where $\mathnormal{y}$ represents the facial feature category, $\mathnormal{y} \in \{\textnormal{EB}, \textnormal{HP}, \textnormal{EAR}, \ldots\}$.  $\mathbf{f}^{\textnormal{S}}_{\textnormal{G}}$ represents the vector of selected global features. Finally, the score obtained from the SVM is denoted as ${\mathnormal{s}_{\textnormal{G}}}$.}
\label{Block_diagram_Features_Fusion}
\end{figure*}

\subsection{Attention Level Estimation based on Facial Features}  \label{sec:Protocol}

Based on the facial features presented in previous sections, we propose a binary classifier to estimate periods of high or low attention.  

The attention levels in the mEBAL2 dataset range from $0$ to $100$; however, for our study, we performed binary classification (high, low). Additionally, the attention levels vary for each student. To address these aspects, we followed the protocol proposed by ALEBk and MATT~\cite{daza2021alebk, daza2023matt}, where two thresholds were defined for high and low attention periods segmentation: high attention (attention higher than a threshold $\tau_H$) and low attention (attention lower than a threshold $\tau_L$). In our case, the thresholds were obtained through the probability density function (PDF) of the attention levels from the $60$ students (see Fig. \ref{fig:Dist_Y}). Specifically, we considered low attention as the values below the $10$th percentile ($\tau_L$) and high attention as the values above the $90$th percentile ($\tau_H$), as these percentiles have been shown in previous works to be separable in high and low attention.

The attention levels from the EEG band are provided every second ($1$ Hz). However, our approach focuses on longer temporal windows to gather enough behavioral and physiological features that can effectively classify attention. Specifically, here we study three different sliding windows of $30$s, $60$s, and $120$s. This means that attention was estimated every second, based on the characteristics extracted from the frame sequence within the time window of size $W_l$ seconds.


We then calculated the band attention level per window (reducing the impact of possible errors and obtaining a more accurate value of the captured attention by the band) and assigned a high or low label. 
After obtaining the labels, we analyzed the video sessions using all modules.  For each facial feature category, we generated two vectors of both local and global features for the applied windows.

We trained two Support Vector Machine (SVM) binary classifiers for each facial feature category, one using local and other using global features as described in Section~\ref{sec:Features} (see Fig.~\ref{Block_diagram}).


All SVMs were trained with a linear kernel, employing a squared L$2$ penalty with a regularization hyper-parameter $C$ ranging from $1e^{-8}$ to $1e^{2}$ with steps in powers of $10$. Additionally, a tolerance of $1e^{-3}$ is set for the stopping criterion. It is important to mention that this work also evaluated the performance of RBF kernel SVM and Random Forest. However, the differences in the performance of the three proposed algorithms were marginal. For greater clarity, the paper only presents the results of the linear SVM classifiers.

\begin{figure*}[t]
    \centering
    \includegraphics[width=\textwidth]{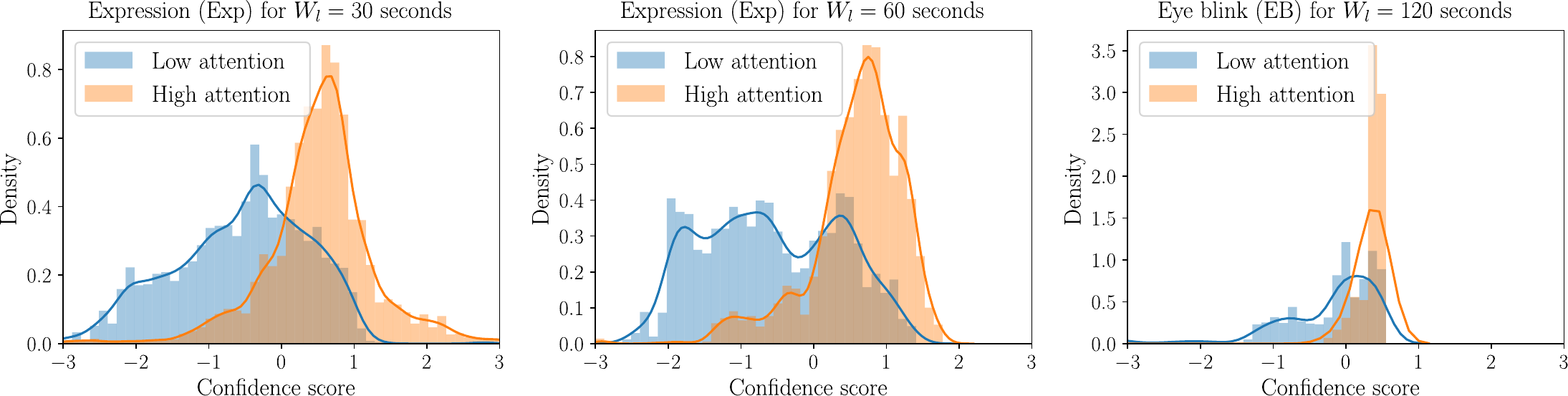} 
    \caption{Probability density distributions of the confidence scores obtained by our attention estimation systems for the best approach in each of the three time windows considered (from left to right: 30s, 60s, and 120s) using local features. In order to simplify the performance analysis/comparison, our experimental discussion is focused on binary classification into low/high attention using the score threshold that maximizes classification accuracy.}
    \label{fig:LocalMonomodal}
\end{figure*}
To obtain the multimodal approach, we applied score level fusion 
with different combinations of the monomodal attention level estimation classifiers, therefore, we sorted out our systems into unimodal and multimodal attention level estimation. The training process works as follows:

\textbf{Unimodal attention level estimation.} 
\textit{i)} Each frame is processed through the $5$ facial analysis modules described in section \ref{sec:FaceAnalysisModule}. \textit{ii)} Output features $\textrm{\textbf{f}}_{x,y}$ are averaged for each second of video $\bar{\textrm{\textbf{f}}}_{x,y}$.  \textit{iii)} A vector  $\textbf{f}^{y}_{\mathrm{L}}$ for local and $\textrm{\textbf{f}}^{y}_{\mathrm{G}}$ for global features are obtained for the time window at hand. The extraction process follows the steps described in the previous section \ref{sec:Features}. Finally, we have the following vectors $\{\textbf{f}_{\mathrm{L}}^{\mathrm{EB}},\textbf{f}_{\mathrm{L}}^{\mathrm{EAR}},\textbf{f}_{\mathrm{L}}^{\mathrm{NS}},\textbf{f}_{\mathrm{L}}^{\mathrm{HS}},\textbf{f}_{\mathrm{L}}^{\mathrm{HP}}, \textbf{f}_{\mathrm{L}}^{\mathrm{H}},
\textbf{f}_{\mathrm{L}}^{\mathrm{Exp}}\}$ and $\{
\textbf{f}_{\mathrm{G}}^{\mathrm{EB}},\textbf{f}_{\mathrm{G}}^{\mathrm{EAR}},\textbf{f}_{\mathrm{G}}^{\mathrm{HS}},\textbf{f}_{\mathrm{G}}^{\mathrm{NS}},\textbf{f}_{\mathrm{G}}^{\mathrm{HP}}, \textbf{f}_{\mathrm{G}}^{\mathrm{H}},  \textbf{f}_{\mathrm{G}}^{\mathrm{Exp}}\}$.
\textit{iv)} Two SVMs for each facial feature category are trained to classify between high and low attention,  one using local features  $\textbf{f}^{y}_{\mathrm{L}}$ and the other using global features $\textrm{\textbf{f}}^{y}_{\mathrm{G}}$ as input. The scores for local features are denoted as ${s_{\mathrm{L}}^y}$, which include $\{s_{\mathrm{L}}^{\mathrm{EB}}, s_{\mathrm{L}}^{\mathrm{EAR}}, s_{\mathrm{L}}^{\mathrm{NS}},s_{\mathrm{L}}^{\mathrm{HS}}, s_{\mathrm{L}}^{\mathrm{HP}}, s_{\mathrm{L}}^{\mathrm{H}}, s_{\mathrm{L}}^{\mathrm{Exp}}\}$ and for global features as ${s_{\mathrm{G}}^y}$, which include $\{s_{\mathrm{G}}^{\mathrm{EB}}, s_{\mathrm{G}}^{\mathrm{EAR}}, s_{\mathrm{G}}^{\mathrm{NS}},s_{\mathrm{G}}^{\mathrm{HS}}, s_{\mathrm{G}}^{\mathrm{HP}}, s_{\mathrm{G}}^{\mathrm{H}}, s_{\mathrm{G}}^{\mathrm{Exp}}\}$.


\textbf{Multimodal attention level estimation}. 
The proposed multimodal systems involve combining unimodal facial analysis systems based on either local or global features.
The scores from previously trained unimodal facial analysis are combined using different strategies: \textit{i)} a score sum strategy and \textit{ii)} training a simple neural network with two hidden layers. The architecture consists of dense layers with ReLU activation. The first hidden layer has 16 units and processes the input, which includes 7 scores,  each corresponding to the output from the SVM binary classifiers for individual facial feature categories. This is followed by another dense layer with 8 units, and an output layer with one unit (sigmoid activation). A dropout of 0.5 is employed.
For both fusion strategies, the process  was carried out individually for local features ${s_{\mathrm{L}}^y}$ and for global features ${s_{\mathrm{G}}^y}$, obtaining two combined scores $s^{\mathrm{F}}_{\mathrm{L}}$ or $s^{\mathrm{F}}_{\mathrm{G}}$.
Finally, these scores were compared with a threshold $\tau$ to determine the attention level (high or low).


We also propose another multimodal system for global features, based on feature selection and fusion using a single SVM classifier (see Fig.~\ref{Block_diagram_Features_Fusion}). The protocol is the same as previously explained; however, instead of training an SVM for each facial feature category, we perform a feature selection and fusion inspired by the work of Leng et al. \cite{leng2017dual}. We merged  global features into a vector $\{\textbf{f}_{\mathrm{G}}^{\mathrm{EB}}, \textbf{f}_{\mathrm{G}}^{\mathrm{EAR}}, \textbf{f}_{\mathrm{G}}^{\mathrm{HS}}, \textbf{f}_{\mathrm{G}}^{\mathrm{NS}}, \textbf{f}_{\mathrm{G}}^{\mathrm{HP}}, \textbf{f}_{\mathrm{G}}^{\mathrm{H}}, \textbf{f}_{\mathrm{G}}^{\mathrm{Exp}}\}$
and calculate the Discrimination Power (DP), which is a measure based on the inter-class and intra-class variation $\mathrm{DP} = \frac{\sigma_{\text{inter}}^2}{\sigma_{\text{intra}}^2}
$. Finally, we select the features that are in the top 90th percentile of DP. This new vector ${\textbf{f}_{\mathrm{G}}^{\mathrm{S}}}$ is used as input to train a single SVM to classify between high and low attention.

\section{EXPERIMENTS AND RESULTS}

\subsection{Experimental Protocol}

We follow the protocol proposed in ALEBK~\cite{daza2021alebk} to classify between high and low attention levels, as detailed in the previous section \ref{sec:Protocol}.  In total, we obtain $10376$, $8309$, and $5605$ periods for time windows $W_l$ of $30$, $60$, and $120$ seconds, respectively, from all $60$ students in the database. The samples are evenly distributed between low and high attention levels.



We employ the leave-one-out cross-validation protocol, where one user is left out for testing, and the remaining ones are used for training and this process is repeated with all users. The decision threshold is chosen at the point where the classification accuracy is maximized. 

\begin{table}[t!]
    \centering
    \caption{Attention estimation Accuracy (Acc in \%) using the mEBAL2 database for the proposed unimodal approaches with local features. We set the value of $\tau_L$ at $10\%$ and $\tau_H$ at $90\%$. The values highlighted in black indicate the best module for each time window ($30$s, $60$s, $120$s)}
    \label{tab:NewResults1_LF}
    \resizebox{\columnwidth}{!}{%
        \begin{tabular}{lccc}
            \noalign{\hrule height 1pt}
              \multirow{2}{*}{\textbf{Module}} &
              \multicolumn{1}{c}{$W_l$ \textbf{: 30 Seconds}} &
              \multicolumn{1}{c}{$W_l$ \textbf{: 60 Seconds}} &
              \multicolumn{1}{c}{$W_l$ \textbf{: 120 Seconds}} \\
             &
              \multicolumn{1}{c}{\textbf{Acc}}&
              \multicolumn{1}{c}{\textbf{Acc}} &
              \multicolumn{1}{c}{\textbf{Acc}} \\
            \noalign{\hrule height 1pt}
            \textbf{Landmark (EAR)}      & 68.52   & 69.84    & 75.54    \\
            \textbf{EyeBlink (EB)}      & 70.54      & 73.91    & \textbf{79.16}  \\
            \textbf{Expression (Exp)}   & \textbf{76.66}     & \textbf{77.28}   & 79.11    \\
            \textbf{Head Pose (HP)}      & 57.21     & 60.61   & 65.23  \\
            \textbf{Landmark (HS)}            & 61.66      & 62.78  & 65.94   \\
            \textbf{Landmark (NS)}           & 62.33     & 62.20   & 57.22    \\
            \textbf{Heart Rate (H)}       & 50.03     & 50.02    & 54.83   \\
            \noalign{\hrule height 1pt}
            \multicolumn{4}{l}{$W_{l}$: Window length (in seconds).}                       
        \end{tabular}%
    }
\end{table}

\subsection{Unimodal Experiments}

We initially divided the experiments into local and global features.


\subsubsection{Local Features}

Table~\ref{tab:NewResults1_LF} displays the results for each facial analysis module in terms of attention estimation Accuracy (Acc in \%) for all time windows ($30$s, $60$s, $120$s). Fig.~\ref{fig:LocalMonomodal}  shows the probability density distributions of the scores obtained for the best method in each window.

\begin{figure*}[t]
    \centering
    \includegraphics[width=\textwidth]{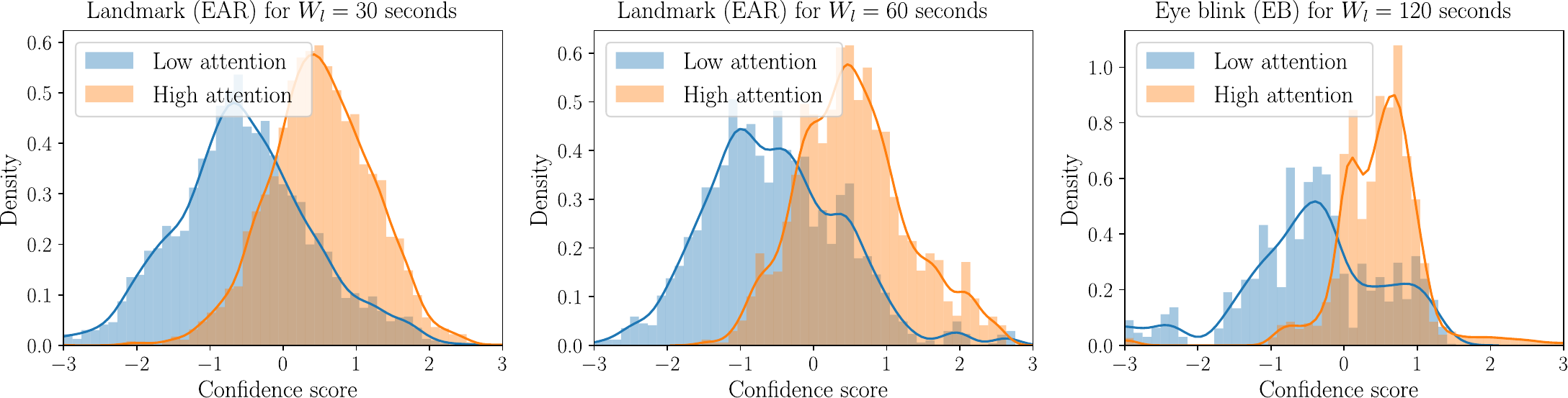} 
    \caption{Probability density distributions of the confidence scores obtained by our attention estimation systems for the best approach in each of the three time windows considered (from left to right: 30s, 60s, and 120s) using global features. In order to simplify the performance analysis/comparison, our experimental discussion is focused on binary classification into low/high attention using the score threshold that maximizes classification accuracy.}
    \label{fig:GlobalMonomodal}
\end{figure*}

The results show that the EyeBlink (EB) and Facial Expression (Exp) modules achieve the highest accuracy with better separability between distributions for all time frames. We noticed that in the $30$s and $60$s windows, the Exp module performs the best with an accuracy of $76.66\%$ and $77.28\%$, respectively. However, in the 120s window, the EB module shows a slight improvement over Exp, achieving an accuracy of $79.16\%$. The third module (feature category) with the best results is the EAR feature category, which reinforces previous findings on the importance of the eye state and facial expressions in attention estimation~\cite{daza2020mebal,daza2021alebk,bagley1979effect, holland1972blinking}.

The worst results are obtained from the Heart Rate (H) module. This suggests that, the variations in Heart Rate do not present a high correlation with  attention levels in this database.

The Head Pose (HP) module has the second worst result for 30s and 60s windows; however, even though it is not a clear attention estimation indicator, it shows that there is a relationship with attention levels, making it potentially useful for multimodal approaches. Additionally, it is observed that as the time window increases, the results improve, reaching an accuracy of $65.23\%$. This makes sense, as a larger window allows capturing more significant patterns and trends in the student's behavior and mitigates possible errors from the pose detection module.

The previous modules show an improvement in the accuracy metric when the time window is extended, with an average improvement of around $6.18\%$. This demonstrates that increasing the amount of features and context allows a better classification. The feature categories based on the eye state are particularly relevant, specifically the EB and EAR, where we observe an accuracy improvement of $8.62\%$ and $7.02\%$ respectively. This makes sense because eyeblinks are less frequent in e-learning environments compared to standard behavior~\cite{portello2013blink, abusharha2017changes}. For this reason, a larger window allows the detection of moments with few eyeblinks (high attention) or periods with a higher eyeblink frequency (low attention).

Similar to HP, head-to-camera indicators like Head and Nose Size, HS and NS respectively, are not strongly correlated to attention. Unlike previous modules, these feature categories do not always perform better in the $120$-second window. This makes sense because during e-learning sessions, students can make fast movements to get closer to the screen, fixing their visual attention on a specific point on the screen, indicating strong concentration. 

Fig.~\ref{fig:LocalMonomodal} shows that, in most cases, high attention levels are easier to recognize than the low ones. Low levels tend to have a more spread density distribution, making their classification more challenging. This makes sense in the context of the monitoring carried out in mEBAL2~\cite{daza2024mebal2}, where students are typically focused with moments of high attention during short time tasks.

\subsubsection{Global Features}

\begin{table}[t!]
    \centering
    \caption{Attention estimation Accuracy (Acc in \%) using the mEBAL2 database for the proposed unimodal approaches with global features. We set the value of $\tau_L$ at $10\%$ and $\tau_H$ at $90\%$. The values highlighted in black indicate the best module for each time window ($30$s, $60$s, $120$s)}
    \label{tab:NewResults1_GF}
    \resizebox{\columnwidth}{!}{%
        \begin{tabular}{lccc}
            \noalign{\hrule height 1pt}
            \multirow{2}{*}{\textbf{Module}} & \multicolumn{1}{c}{$W_l$\textbf{: 30 Seconds}} & \multicolumn{1}{c}{$W_l$\textbf{: 60 Seconds}} & \multicolumn{1}{c}{$W_l$\textbf{: 120 Seconds}} \\
                & \textbf{Acc}   & \textbf{Acc}     &   \textbf{Acc}  \\
            \noalign{\hrule height 1pt}
            \textbf{Landmark (EAR)}     & \textbf{75.94}   & \textbf{75.87}   & 75.99     \\
            \textbf{EyeBlink (EB)}    & 73.03     & 74.41       & \textbf{80.64}           \\
            \textbf{Expression (Exp)}   & 73.05       & 74.46       & 78.39          \\
            \textbf{Head Pose (HP)}    & 63.62       & 59.78      & 59.52           \\
            \textbf{Landmark (HS)}      & 64.12      & 62.09       & 53.79           \\
            \textbf{Landmark (NS)}    & 63.07      & 62.17        & 56.32            \\
            \textbf{Heart Rate (H)}     & 52.35      & 55.84       & 59.86            \\
            \noalign{\hrule height 1pt}
            \multicolumn{4}{l}{$W_{l}$: Window length (in seconds).}                            \\
        \end{tabular}%
    }
\end{table}

We conducted the same experiments as in the previous section with the global features to understand if they are more effective in the SVM-based classification and how they impact each module. This analysis aims to assess whether the global features can provide additional discriminating information to improve the accuracy of attention estimation compared to the local features.

Table~\ref{tab:NewResults1_GF} shows the results for each module in different time windows ($30$s, $60$s, $120$s). Similar to the previous case, the probability density distributions of the  scores for the best method in each window are shown in Fig.~\ref{fig:GlobalMonomodal}. 

The EAR feature category achieves the best results in the $30$s and $60$s windows, achieving a maximum accuracy of $75.94\%$ and $75.87\%$, respectively. We can observe significant improvements in the results of this module in comparison to local features, achieving an accuracy improvement of $7.42\%$ and $6.03\%$, respectively. Once again, the top three feature categories with the best results are EAR, EB, and Exp.

\begin{table*}[t]
\caption{Accuracy results (Acc in \%) for attention estimation in multimodal systems based on local features, showing the best combinations for score sum fusion. The first row provides the best unimodal module for the selected time window. The last row displays the results achieved by score fusion via neural network.  The values highlighted in black indicate the best feature categories and the fusion strategy with the best accuracy for each time window ($30$s, $60$s, $120$s)}
\label{tab:NewResults2_LF}
\resizebox{\textwidth}{!}{%
\begin{tabular}{l}
    \begin{tabular}{lc}
    \noalign{\hrule height 1pt}
    \multicolumn{2}{c}{$W_l$\textbf{ : 30 Seconds}}                                   \\
    \textbf{Feature Categories}           & \textbf{Acc} \\
    \noalign{\hrule height 1pt}
    Exp                        & 76.66         \\
    EB, Exp                    & \textbf{77.25}            \\
    EB, Exp, HS                & 73.95              \\
    Exp, EAR, HP, HS           & 73.00               \\
    EB, Exp, EAR, HP, HS       & 73.18                \\
    EB, Exp, EAR, HP, H, HS   & 70.46               \\
    All    Modules                    & 68.67              \\
            
    Neural Network Fusion   & \textbf{84.25}              \\
    \noalign{\hrule height 1pt}
    \end{tabular}
    \begin{tabular}{lc}
    \noalign{\hrule height 1pt}
    \multicolumn{2}{c}{$W_l$\textbf{ : 60 Seconds}}                                   \\
    \textbf{Feature Categories}           & \textbf{Acc}  \\
    \noalign{\hrule height 1pt}
    Exp                      & 77.28                \\
    EB, Exp                  & \textbf{77.65}            \\
    EB, Exp, HP              & 75.92                \\
    EB, Exp, EAR, HP         & 76.48               \\
    EB, Exp, EAR, HP, HS     & 75.11              \\
    EB, Exp, EAR, HP, HS, NS & 72.10                \\
    All  Modules                    & 70.01            \\
    Neural Network Fusion   & \textbf{85.87} 
         \\
    \noalign{\hrule height 1pt}
    \end{tabular}
    \begin{tabular}{lc}
    \noalign{\hrule height 1pt}
    \multicolumn{2}{c}{$W_l$\textbf{ : 120 Seconds}}                                  \\
    \textbf{Feature Categories}           & \textbf{Acc}  \\
    \noalign{\hrule height 1pt}
    EB                         & 79.16               \\
    EB, Exp                     & \textbf{80.52}          \\
    EB, Exp, HP                 & 80.32               \\
    EB, Exp, EAR, H            & 79.95                \\
    EB, Exp, EAR, HP, H        & 78.18                \\
    EB, Exp, EAR, HP, H, HS    & 76.88                \\
    All Modules                        & 76.25            \\
     Neural Network Fusion        &  \textbf{85.92}           \\
    \noalign{\hrule height 1pt}
    \end{tabular}%
    \\
    $W_{l}$: Window length (in seconds).
\end{tabular}%
}

\end{table*}

In the case of the EB module, we can see improvements in all three windows, but a notable difference in the 120-second window. The accuracy in this case reaches $80.64\%$, which is the highest obtained value. 

The Exp module shows a decrease in accuracy results compared to the local features in all three windows, with differences of $3.61\%$ for the 30s window, $2.82\%$ for the 60s window, and $0.72\%$ for the 120s window, noticing an error reduction as the window size increases. 

The Heart Rate module remains an unreliable indicator for attention estimation, as its classification is almost random in the considered time windows.
The other features categories, user distance and head pose, exhibit similar behavior, showing slight improvements in the first window and deterioration in the subsequent ones when compared to local features. By themselves do not serve as a clear indicator for attention estimation. However, as we will see later, the information provided by these features categories might be valuable in multimodal systems.

Results show that our best unimodal models improve their performance as the temporal window increases up to $120$ seconds. The same trend is observed with local features, highlighting the importance of considering a longer time period to capture significant patterns and trends in the students' behavior. This finding supports the notion that certain discriminating features may become clearer and more effective in attention estimation when analyzing a larger temporal context. By expanding the window, we allow the modules to detect and utilize more relevant information for classification, resulting in an enhanced ability to distinguish between high and low attention levels with greater accuracy.
As we can see, some modules were significantly improved using global features, such as EAR feature category. Additionally, the size of the temporal windows has a notable impact on the results. Global features achieve the highest accuracy value of $80.64\%$ for the EB module.

Figure~\ref{fig:GlobalMonomodal} also shows that detecting low attention levels can be more challenging than detecting higher ones, because the low attention score distribution is more spread than the high attention one. Although this difference is not as clear in global features as it is in local ones, it is particularly evident in the 120s window.

\subsection{Multimodal Experiments}


\subsubsection{Local Features}

Table~\ref{tab:NewResults2_LF} displays the results from the best combinations of unimodal for the score sum strategy and the score fusion results using a neural network. 

The best results in the 30s window for score sum are achieved combining the EB+Exp modules, with an accuracy of $77.25\%$. Compared to the Exp module, which is the best unimodal module, there is a slight improvement of $0.60\%$. We observe that the combination with other modules worsens the results compared to the Exp module. However, the score fusion using a neural network achieves the best performance with $84.25\%$, marking a significant improvement over the EB+Exp module combination by 7\%, demonstrating the potential of neural networks for score fusion \cite{fierrez2018multiple}.


The same pattern occurs in the 60s window, making the EB+Exp combination the best for score sum, showing a slight improvement over the Exp module alone. The other combinations result in a worse performance. Once again, in the 60s window, the Neural Network Fusion (NNF) achieves the best results, even surpassing those in the 30s window. NNF outperforms both the unimodal system, with a significant improvement of 8.59\%, and the top-performing EB+Exp combination by 8.22\%.

In the 120s window the best unimodal system is EB, which is slightly surpassed by three different fusions: EB+Exp, EB+Exp+HP, and EB+Exp+EAR+H. The best maximum accuracy is achieved by EB+Exp with $80.52\%$ for score sum. Once again, NNF outperforms the score sum, achieving an accuracy of $85.92\%$. This demonstrates that increasing the temporal window improves the system combination accuracy, as expected because a broader temporal context facilitates the integration of longer and more complex temporal patterns in the data, resulting in better discrimination. Furthermore, these results show that score-level fusion with neural networks is more effective than score sum for local features in the mEBAL2 database.


\begin{table*}[t]
\caption{Accuracy results (Acc in \%) for attention estimation in multimodal systems based on global features, showing the best combinations for score sum fusion. The first row provides the best unimodal module for the selected time window. The last row displays the results achieved by score fusion via neural network.  The values highlighted in black indicate the best feature categories and the fusion strategy with the best accuracy for each time window ($30$s, $60$s, $120$s)}
\label{tab:NewResults2_GF}
\resizebox{\textwidth}{!}{%
\begin{tabular}{l}
    \begin{tabular}{lc}
    \noalign{\hrule height 1pt}
    \multicolumn{2}{c}{$W_l$\textbf{ : 30 Seconds}}                                 \\
    \textbf{Feature Categories}         & \textbf{Acc} \\
    \noalign{\hrule height 1pt}
    EAR                      & 75.94                                      \\
    EAR, Exp                 & 76.77                                       \\
    EB, EAR, HS              & 77.11                                         \\
    EB, Exp, EAR, HP         & 77.35                                       \\
    EB, Exp, EAR, HP, HS     & \textbf{77.39}                               \\
    EB, Exp, EAR, HP, HS, H & 77.31                                       \\
    All   Modules                    & 76.73                                     \\
       Neural Network Fusion        &  \textbf{79.26}           \\
    \noalign{\hrule height 1pt}
    \end{tabular}
    \begin{tabular}{lc}
    \noalign{\hrule height 1pt}
    \multicolumn{2}{c}{$W_l$\textbf{ : 60 Seconds}}                                 \\
    \textbf{Feature Categories}         & \textbf{Acc}  \\
    \noalign{\hrule height 1pt}
    EAR                      & 75.87                                    \\
    EB, EAR                  & 79.17                                     \\
    EB, EAR, NS              & 78.90                                      \\
    EB, Exp, EAR, NS         & 79.01                                 \\
    EB, Exp, EAR, HP, NS     & \textbf{79.23}                                \\
    EB, Exp, EAR, HP, H, NS & 78.32                                      \\
    All   Modules                  & 77.05                                 \\
        Neural Network Fusion        &  79.17
         \\
    \noalign{\hrule height 1pt}
    \end{tabular}
    \begin{tabular}{lc}
    \noalign{\hrule height 1pt}
    \multicolumn{2}{c}{$W_l$\textbf{ : 120 Seconds}}                                \\
    \textbf{Feature Categories}         & \textbf{Acc}  \\
    \noalign{\hrule height 1pt}
    EB                       & 80.64                                  \\
    EB, Exp                  & \textbf{83.34}                                 \\
    EB, Exp, EAR             & 80.95                                      \\
    EB, Exp, EAR, NS         & 79.80                                    \\
    EB, Exp, EAR, H, NS     & 77.23                                      \\
    EB, Exp, EAR, H, HS, NS & 76.16                                        \\
    All  Modules                   & 73.10                                    \\
            Neural Network Fusion        &  74.01  
         \\
    \noalign{\hrule height 1pt}
    \end{tabular}
    \\
    $W_{l}$: Window length (in seconds).
\end{tabular}%
}
\end{table*}

The EB and Exp unimodal modules are the most effective in attention estimation, appearing in all combinations that improved results. Additionally, the best values are consistently obtained in the 120s window, having the most potential for improvement due to the wider amount of information. This highlights the challenge of attention estimation through image processing, requiring longer windows to capture relevant behavioral and physiological processes.

\begin{figure}[t]
    \centering
    \includegraphics[width=\columnwidth]{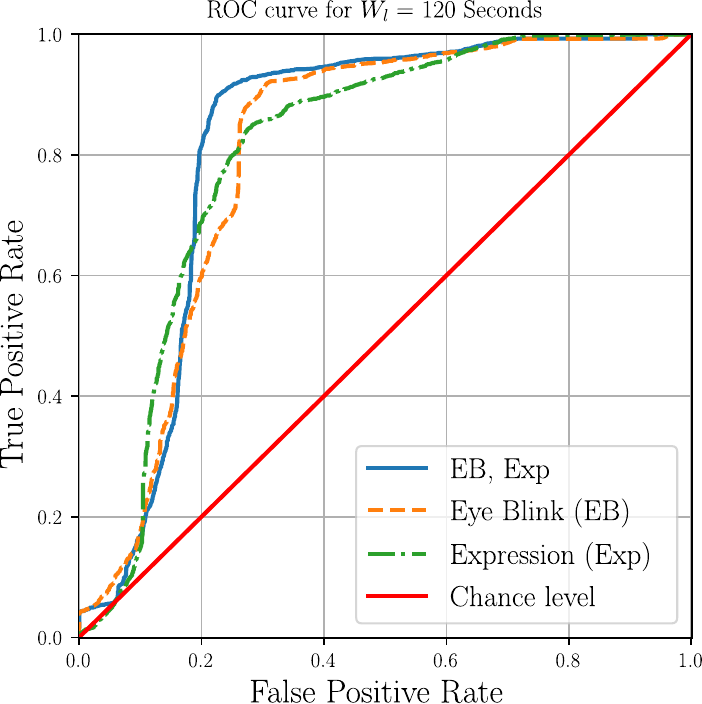} 
    \caption{Receiver Operating Characteristic curve (ROC) obtained for the most accurate multimodal approach using global features (this occurs in the 120s window), shown with a blue line and for each of the monomodal systems that are part of this combination.}
    \label{fig:GF_Best}
\end{figure}

\subsubsection{Global Features}

Table~\ref{tab:NewResults2_GF} presents the results of the best combinations with global features for score sum and neural network fusion. Fig.~\ref{fig:GF_Best}  shows the ROC curve for the best multimodal system, along with the results of the individual monomodal systems that compose it.

With global features, more effective combinations are achieved for score sum and lower performances are obtained for NNF. In the 30s window, the best unimodal result is $75.94\%$, and all combinations shown in the table (combining with the EAR feature category) outperform it, including the combination of all modules. The best one is EB+Exp+EAR+HP+HS with $77.39\%$, representing an average improvement of $1.65\%$ in accuracy. Furthermore, this combination slightly outperforms ($0.14\%$) the best result with local features, which was EB+Exp in the 30s window. However, EB+Exp requires only two modules, making it faster and more practical.
NNF achieves better performance than the best combination by score sum, with a slight improvement of 1.87\%. However, its performance is inferior to the results obtained for local features in all windows.

In the 60s window, we observe a similar pattern for the score sum. The unimodal EAR feature category achieves an accuracy of $75.87\%$ and all the combinations outperform it when combined with the same feature category. The best combination is EB+Exp+EAR+HP+NS, similar to the previous one, with an accuracy of $79.23\%$. This combination shows a significant improvement of $3.36\%$ in accuracy. 
This highlights that the user distance and pose feature categories contain valuable information in multimodal systems, especially in short-duration windows. Furthermore, the best combination of global features surpasses the results of the best combination with local features, EB+Exp, by $1.58\%$ in accuracy, demonstrating a considerable improvement.
However, for NNF, the results are similar to those in the 30s window. For the first time, this method is inferior to the best combination for score sum, though almost equal. Once again, the results obtained by NNF for global features are inferior to those achieved with local features.

For the 120s window, we found out that the combination of EB+Exp significantly outperformed the best unimodal approach, which was EB. On the other hand, EB+Exp achieves $83.34\%$ resulting in a remarkable improvement of $2.7\%$ in accuracy. Furthermore, global features also surpass the best combination with local features for score sum by $2.82\%$,  showing that global features continue to achieve a better combination of modules, especially in the 120s window where the best results are obtained. Additionally, in this window, the best combinations for both local and global features are EB+Exp, indicating that, under similar conditions, the most effective system is achieved with global features. Additionally, this also highlights the importance of facial units and EyeBlink in attention estimation. NNF achieves its lowest performance with an accuracy of 74.01\%, which is 9.33\% lower than the best combination of EB+Exp. It has been observed that score fusion using global features has worse generalization compared to local features

Figure~\ref{fig:GF_Best} presents the ROC curve for the unimodal and multimodal approaches based on global features and $120$s window (best approaches). The curve shows significant improvement when combining Facial Expressions and EyeBlink.


\subsubsection{Global Features: Selection and Feature Level Fusion}

\begin{table}[th]
    \caption{Best results of accuracy for global features in unimodal system, score level fusion, and feature level fusion  with the best accuracy for each time window (30s, 60s, 120s)
    }
    \label{tab:Feature level fusion}
    \resizebox{\columnwidth}{!}{%
    \begin{tabular}{lccc}
        \noalign{\hrule height 1pt}
         & $W_l$\textbf{ : 30s} & $W_l$\textbf{ : 60s} & $W_l$\textbf{ : 120s} \\\textbf{Methods}
                         & \textbf{Acc} & \textbf{Acc} & \textbf{Acc} \\
        \noalign{\hrule height 1pt}
        Unimodal   System  &      75.94     &    75.87        & 80.64       \\
        Score Level Fusion             &      79.26     &    79.23        & 83.34        \\
        Feature Level Fusion              & 76.48           &  77.79         & 81.48       \\
        \noalign{\hrule height 1pt}
    \end{tabular}%
    }
\end{table}

Fig.~\ref{Block_diagram_Features_Fusion}  shows the proposed architecture for feature selection and fusion (for more details, see Section \ref{sec:Protocol}). Table \ref{tab:Feature level fusion} presents the best results achieved for unimodal systems, score level fusion, and feature level fusion using global features.

The results show that the proposed architecture for feature selection and fusion outperforms unimodal systems in all windows. However, it produces inferior results compared to the top-performing multimodal systems achieved through score fusion. However, our feature level fusion architecture only utilizes 10\% of the global features (reducing from 728 features to 73) and uses only an SVM to obtain a direct score without the need to train a neural network, which requires more careful optimization. This demonstrates that the results of this architecture are highly competitive.

\subsection{Experiments: comparison with existing approaches}

We now compare ourselves with three recent state-of-the-art approaches: Peng~\cite{peng2020predicting}, ALEBk~\cite{daza2021alebk}, and MATT~\cite{daza2023matt}. ALEBk and MATT were previously trained and evaluated on the first version of mEBAL~\cite{daza2020mebal}.
To perform the comparison here, we train again all the methods on mEBAL2 with $60$ users under identical conditions, using the same attention classification percentile, and employing the leave-one-out cross-validation protocol.
Table~\ref{tab:summaries} presents a benchmark with the best results obtained in attention estimation on the mEBAL2 dataset~\cite{daza2024mebal2}  by different state-of-the-art approaches, compared to our proposal here: DeepFace-Attention.

\begin{table}[th]
    \caption{Comparison with the state of the art. Attention level estimation results on the mEBAL2 dataset~\cite{daza2024mebal2} including 60 students. Our best approach is compared with Peng~\cite{peng2020predicting}, ALEBk~\cite{daza2021alebk} and MATT~\cite{daza2023matt}. The same training and evaluation protocol is employed for all methods following our experimental protocol.\ * We have adapted the method Peng~\cite{peng2020predicting} for classifying between high and low attention levels. This method was designed to work with global features extracted from the head pose module and the facial landmark module.\ **We have adapted the methods proposed in \cite{daza2021alebk} and \cite{daza2023matt} incorporating the global and local features proposed in this work. The results obtained for the 120s time frame are shown, which exhibited the highest accuracy for the best-performing approaches. 
}
    \label{tab:summaries}
    \resizebox{\columnwidth}{!}{%
    \begin{tabular}{lcc}
        \noalign{\hrule height 1pt}
            & \multicolumn{1}{c}{\textbf{Local Features}}          & \multicolumn{1}{c}{\textbf{Global Features}}         \\
        \textbf{Methods} & \textbf{Acc}  & \textbf{Acc}  \\
        \noalign{\hrule height 1pt}
        Peng~\cite{peng2020predicting}*           & --         & 66.28         \\
        ALEBk** \cite{daza2021alebk}           & 79.16         & 80.64         \\
        MATT** \cite{daza2023matt}            & 80.32         & 74.43         \\
        Proposed: DeepFace-Attention  & \textbf{85.92}         & 83.34         \\
        \noalign{\hrule height 1pt}
    \end{tabular}%
    }
\end{table}

The approach proposed in Peng~\cite{peng2020predicting} is a multimodal system based on global features of head posture and movements of the eyes, head, and mouth (see Section \ref{sec: State of art methods} for more information). This approach was based on a random forest model to estimate attention in 10s windows. We adapted this model to predict high and low attention in 30s, 60s, and 120s windows. The results obtained are inferior compared to the methods ALEBk\cite{daza2021alebk}, MATT\cite{daza2023matt}, and our method. Regarding global features, our method improves the performance by 17.06\% over Peng~\cite{peng2020predicting} and by 19.64\% over our best method based on local features. 

ALEBk~\cite{daza2021alebk} is a monomodal system that estimates attention based on the eyeblink rate per minute. 
An enhanced version of that system was used, incorporating an SVM for high and low attention classification, using local and global features, obtained from the eyeblink detector, rather than simply applying a blink rate per minute threshold.
While ALEBk achieved an accuracy of 74\% for the first version of mEBAL, the improved version obtains 79.16\% when applied to mEBAL2 with 60 users. Furthermore, the results of employing global features have been also evaluated over ALEBk, achieving an accuracy of 80.64\%.

MATT~\cite{daza2023matt} presented unimodal and multimodal approaches to classify between high and low attention levels. Its best-performing approach was the multimodal one, which combined EyeBlink, Head Pose, and Facial Expression, using local features. This method achieved an accuracy of 80.32\% with local features and 74.43\% with global features. Better results are obtained with local features compared to the global ones, in contrast to the outcomes achieved by ALEBk. 

As seen on Table~\ref{tab:summaries}, our approach outperforms previous approaches. Our best multimodal approach is EB+Exp score sum combination for global features and NNF for local features. The best results are achieved with local features, surpassing the latest version of ALEBk by 6.8\%, resulting in a relative reduction in error rates of 32.4\%. Regarding the best version of MATT, corresponding to the use of local features, an improvement of 5.6\%  in accuracy is obtained, leading to a relative reduction in error rates of 28.5\%. 
Our global feature system based on score sum also outperforms state-of-the-art proposals using global features with a relative reduction in error rates of 14\% for ALEBk and 34.8\% for MATT.

Furthermore, our multimodal system based on score sum requires only two modules (EB+Exp), while the MATT approach requires three (EB+Exp+HP), resulting in reduced time and resource usage.


\begin{table}[t!]
    \centering
    \caption{Comparative Inference Times for Attention Estimation using an Intel Core i5-7600 CPU, RAM 32GB of RAM, and a NVIDIA GTX 1080 GPU  with 8GB of VRAM. This table includes inference times of the facial analysis modules and a comparison with state-of-the-art methods. Our best approach is compared with Peng~\cite{peng2020predicting}, ALEBk~\cite{daza2021alebk}, and MATT~\cite{daza2023matt}.}
    \label{tab:combined_inference_comparison}
    {\normalsize
    \begin{tabular}{lc}
    
        \toprule
        \textbf{Modules} & \textbf{Inference speed (ms)} \\
   
             \midrule
        
        Face Detection & 29.76 \\
        EyeBlink & 15.94 \\
        Landmark & 41.45 \\
        Head Pose & 101.57 \\
        Expression & 14.59 \\
        Heart Rate & 5.15 \\
        \midrule
        \textbf{Methods} & \textbf{Inference speed (ms)} \\
        \midrule
        Peng \cite{peng2020predicting} & 172 \\
        ALEBk \cite{daza2021alebk} & 86 \\
        MATT \cite{daza2023matt}  & 202 \\
        Proposed (Global Features) & 100 \\
        Proposed (Local Features) & 208 \\
        \bottomrule
    \end{tabular}
    }
\end{table}

Table~\ref{tab:combined_inference_comparison} presents the results of the average inference speed for each processed frame by different facial modules using an Intel Core i5-7600 CPU,  32GB of RAM, and a NVIDIA GTX 1080 GPU  with 8GB of VRAM, providing a computational comparison of the various modules. 
It also presents the inference times per processed frame for our methods and the state-of-the-art methods: Peng~\cite{peng2020predicting}, ALEBk~\cite{daza2021alebk}, and MATT~\cite{daza2023matt}. The Head Pose module is the slowest, followed by the landmark module. Although the EyeBlink module only takes 15.94 ms, using the landmark module is necessary to identify the eye region, thus the total time of 57 ms.  As we can see in Table~\ref{tab:combined_inference_comparison}, the slowest methods are the systems that utilize the Head Pose, such as Peng~\cite{peng2020predicting}, MATT~\cite{daza2023matt}, and our local features NNF method.
The fastest method is ALEBk~\cite{daza2021alebk}, which only uses the eyeblink and landmark modules. 

Note that our objective in this research was not to enhance the system's speed but rather to evaluate whether deep learning-based facial analysis modules can accurately determine high or low attention levels. For future work, resource-optimized modules can be used to reach real-time operation if needed.



\section{CONCLUSION}

We have presented various approaches to estimate high or low attention levels, applied to a realistic e-learning environment of $60$ students.
State-of-the-art technologies were used, based on deep learning, to perform facial analysis of behavioral features and physiological processes related to attention \cite{daza2023matt, daza2024mebal2, zaletelj2017predicting}. To understand which features are more efficient in attention estimation, we designed unimodal systems based on SVM classification using the following information: eyeblink, heart rate, facial expressions, head pose, and head distance.
We also have investigated the impact of local features and well-known global features on accuracy. Additionally, we examined the effects of temporal windows on attention estimation, with three different options: $30$, $60$, and $120$ seconds.
We proposed multimodal systems for attention estimation, demonstrating their ability to enhance existing methods for attention estimation.

Some interesting findings are as follows: eye state features (EAR, EyeBlink) and facial expressions are the most useful with a clear correlation with attention. We also observed that the best attention estimation systems improved as the time window size increases. Head pose and distance features were not clear indicators of attention; however, in multimodal systems, they provided relevant information for classification. The results of the Heart Rate module, both unimodal and combined, showed that it is not a reliable indicator of attention. Global features were more effective for multimodal systems based on score sum, obtaining the best combination with Eyeblink and Facial Expressions with an accuracy of $83.34\%$.
The best results in this study were achieved with local features using score level fusion through neural network training with an accuracy of $85.92\%$. We also analyzed an architecture based on the selection and fusion of global features, outperforming unimodal systems with slightly less accuracy than our full score fusion, but only necessitating 10\% of the features.

Our best approach, called DeepFace-Attention, have outperformed three state-of-the-art methods: Peng~\cite{peng2020predicting}, ALEBk~\cite{daza2021alebk},  and MATT~\cite{daza2023matt}; 
achieving a significant relative improvement in error reduction of approximately 50.6\% for Peng~\cite{peng2020predicting}, 32.4\% for ALEBk (an enhanced version of the system proposed in ~\cite{daza2021alebk}), and 28.5\% for MATT. 


In the future, we will explore the combination of local and global features during the training process. Moreover, we aim to analyze how attention estimation can be affected when students perform different types of tasks. Additionally, we will explore alternative indicators that have shown a direct relation with attention levels, such as eye pupil size \cite{rafiqi2015pupilware, krejtz2018eye}, gaze tracking \cite{wang2014eye, Navarro2024}, keystroking \cite{2016_IEEEAccess_KBOC_Aythami,morales2016kboc, 2024_Access_KVC_Straga}, among others. Predicting the level of attention within a continuous range is a more challenging task than predicting high or low attention levels, and it is also planned for future work.

\bibliographystyle{IEEEtran} 
\bibliography{biography}
\begin{IEEEbiography}[{\includegraphics[width=1in,height=1.25in,clip,keepaspectratio]{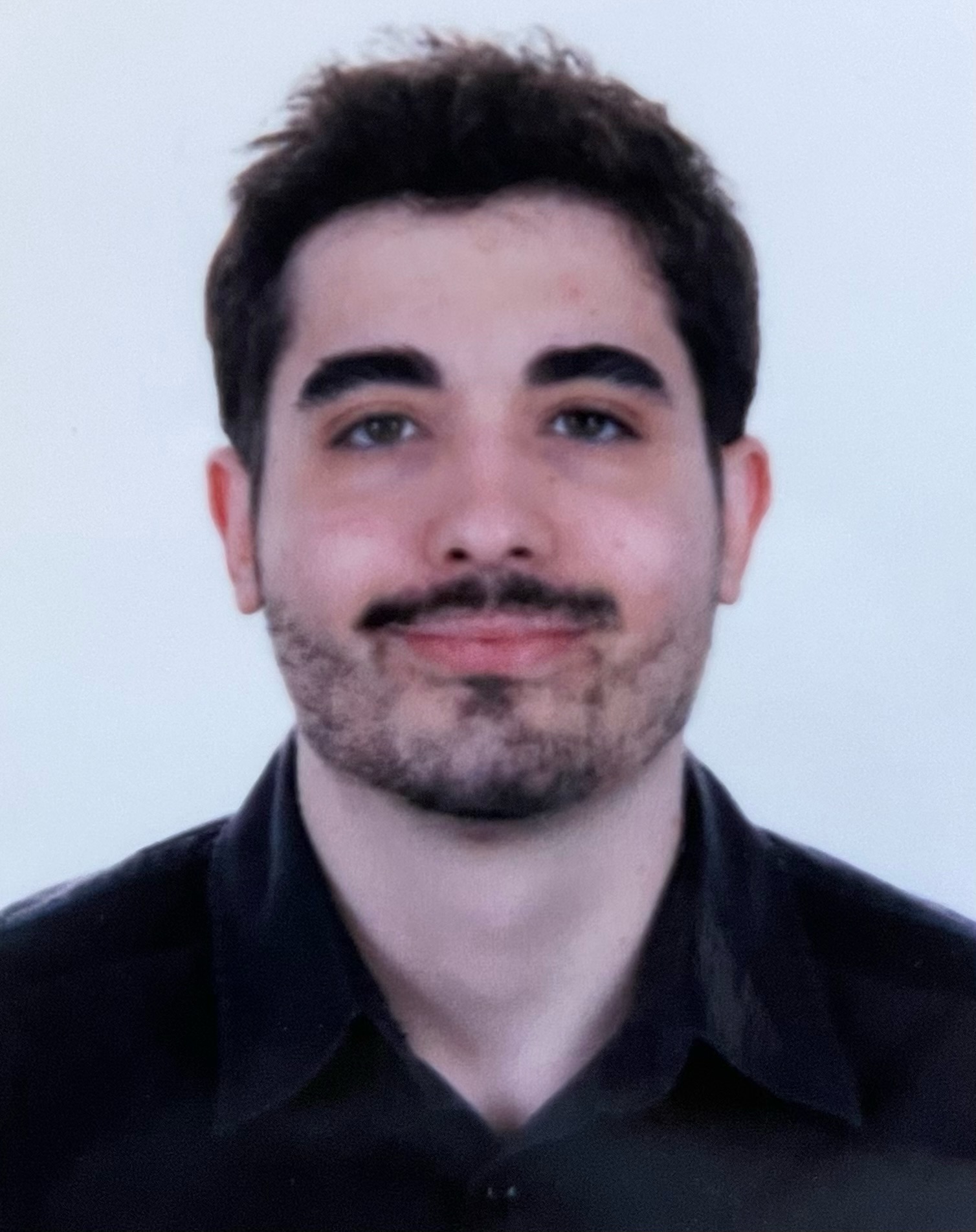}}]{Roberto Daza} received his bachelor’s degree in telecommunications technology engineering in 2016 from Universidad de Granada. He received his M.Sc. degree in Telecommunication Engineering in 2019 from Universidad Oberta de Catalunya and the Universitat Ramon Llull-La Salle. He is pursuing his PhD degree at the Universidad Autónoma de Madrid in the Biometrics and Data Pattern Analytics Lab. His research interests include human-machine interaction, machine learning, deep learning and biometrics signal processing, with an emphasis on e-learning technologies for security and learning improvement. He has received awards from the eMadrid network, AERFAI, Universidad Autónoma de Madrid and doctoral schools all over Madrid. Finally, he has participated in several National and European projects focused on the improvements of e-learning, health technologies and security. 
\end{IEEEbiography}

\begin{IEEEbiography}[{\includegraphics[width=1in,height=1.25in,clip,keepaspectratio]{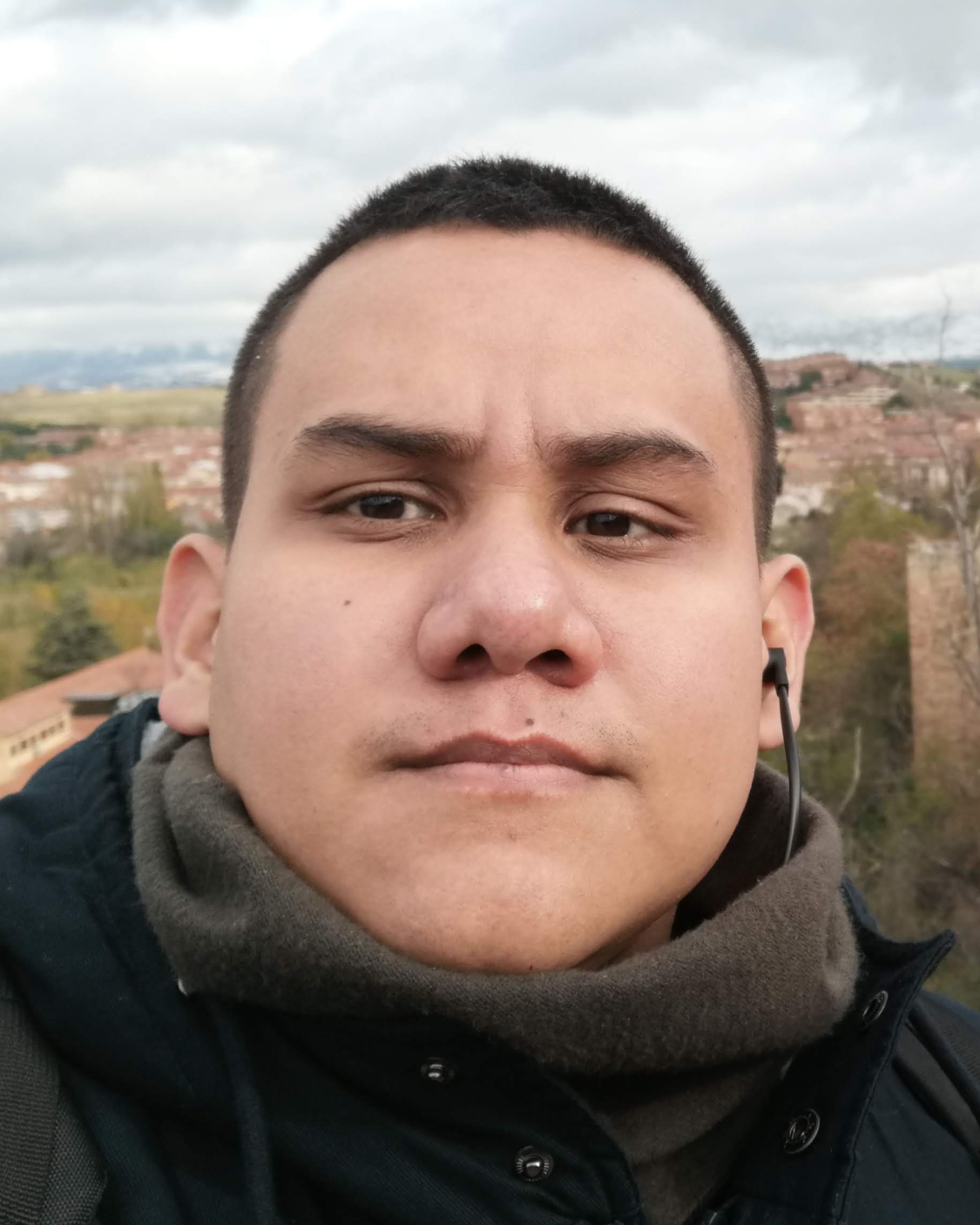}}]{Luis F. Gomez} received his M.Sc. and bachelor's degree in Telecommunications Engineering from Universidad de Antioquia, Medellin, Colombia, in 2018 and 2021. He is a PhD Candidate at the BiDA Lab from the Universidad Autónoma de Madrid. During the last five years, he has performed research activities on signal processing image processing, pattern recognition,  machine learning, and deep learning, with a focus on biometrics signal processing and their health care and security applications with academic and industrial partners.
\end{IEEEbiography}

\begin{IEEEbiography}[{\includegraphics[width=1in,height=1.25in,clip,keepaspectratio]{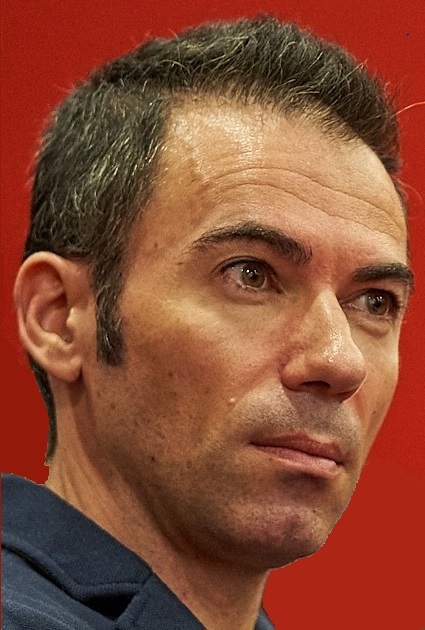}}]{Julian Fierrez} received the M.Sc. and the Ph.D. degrees in telecommunications engineering from Universidad Politecnica de Madrid, Spain, in 2001 and 2006, respectively. Since 2002 he was affiliated as a PhD candidate with the Universidad Politecnica de Madrid, and since 2004 at Universidad Autonoma de Madrid, where he is currently a Full Professor since 2022. From 2007 to 2009 he was a visiting researcher at Michigan State University in USA under a Marie Curie fellowship. Since 2016 he is Associate Editor for Elsevier's Information Fusion and IEEE Trans. on Information Forensics and Security, and since 2018 also for IEEE Trans. on Image Processing. He has been General Chair of the IAPR Iberoamerican Congress on Pattern Recognition (CIARP 2018) and the Iberian Conference on Pattern Recognition and Image Analysis (IbPRIA 2019). He is also recipient of a number of world-class research distinctions, including: EBF European Biometric Industry Award 2006, EURASIP Best PhD Award 2012, Medal in the Young Researcher Awards 2015 by the Spanish Royal Academy of Engineering, and the Miguel Catalan Award. 
\end{IEEEbiography}

\begin{IEEEbiography}[{\includegraphics[width=1in,height=1.25in,clip,keepaspectratio]{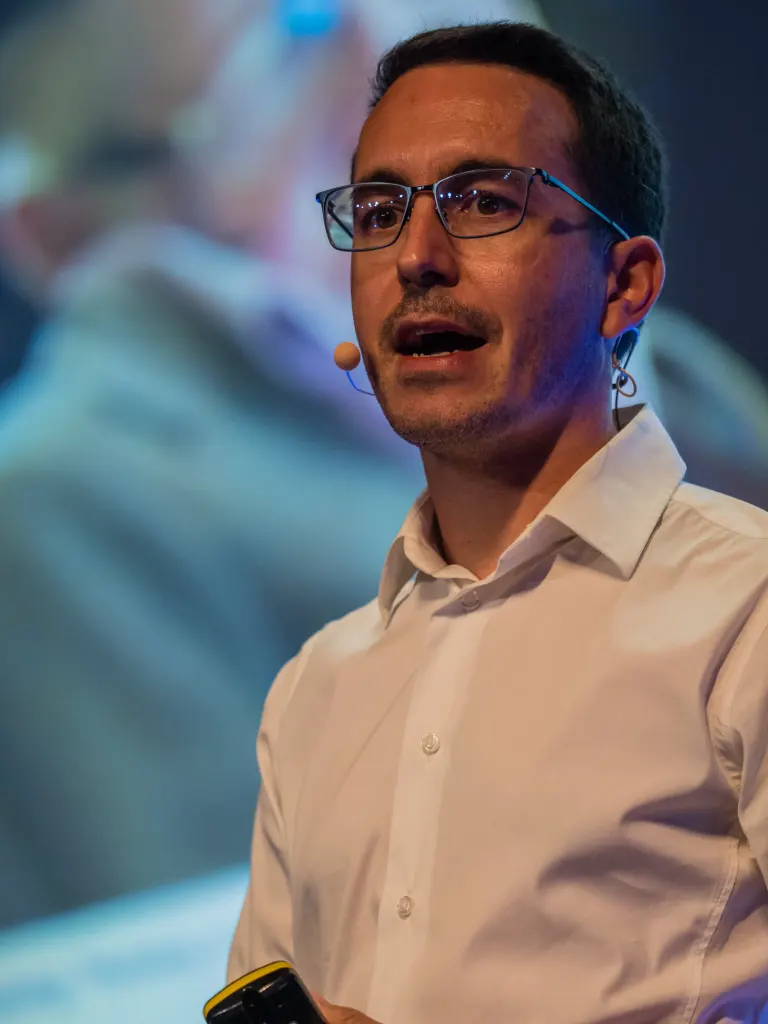}}]{Aythami Morales} received his M.Sc. degree in Electrical Engineering in 2006 from Universidad de Las Palmas de Gran Canaria. He received his Ph.D. degree in Artificial Intelligence from La Universidad de Las Palmas de Gran Canaria in 2011. He performs his research works in the BiDA Lab – Biometric and Data Pattern Analytics Laboratory at Universidad Autónoma de Madrid, where he is currently an Associate Professor (CAM Lecturer Excellence Program). He is member of the ELLIS Society (European Laboratory for Learning and Intelligent Systems). He has performed research stays at the Biometric Research Laboratory at Michigan State University, the Biometric Research Center at Hong Kong Polytechnic University, the Biometric System Laboratory at University of Bologna and Schepens Eye Research Institute (Harvard Medical School). He is author of more than 100 scientific articles published in international journals and conferences, and 2 patents. A. Morales is supported by the Madrid Government (Comunidad de Madrid-Spain) under the Multiannual Agreement with Universidad Autónoma de Madrid in the line of Excellence for the University Teaching Staff in the context of the V PRICIT (Regional Programme of Research and Technological Innovation).
\end{IEEEbiography}

\begin{IEEEbiography}[{\includegraphics[width=1in,height=1.25in,clip,keepaspectratio]{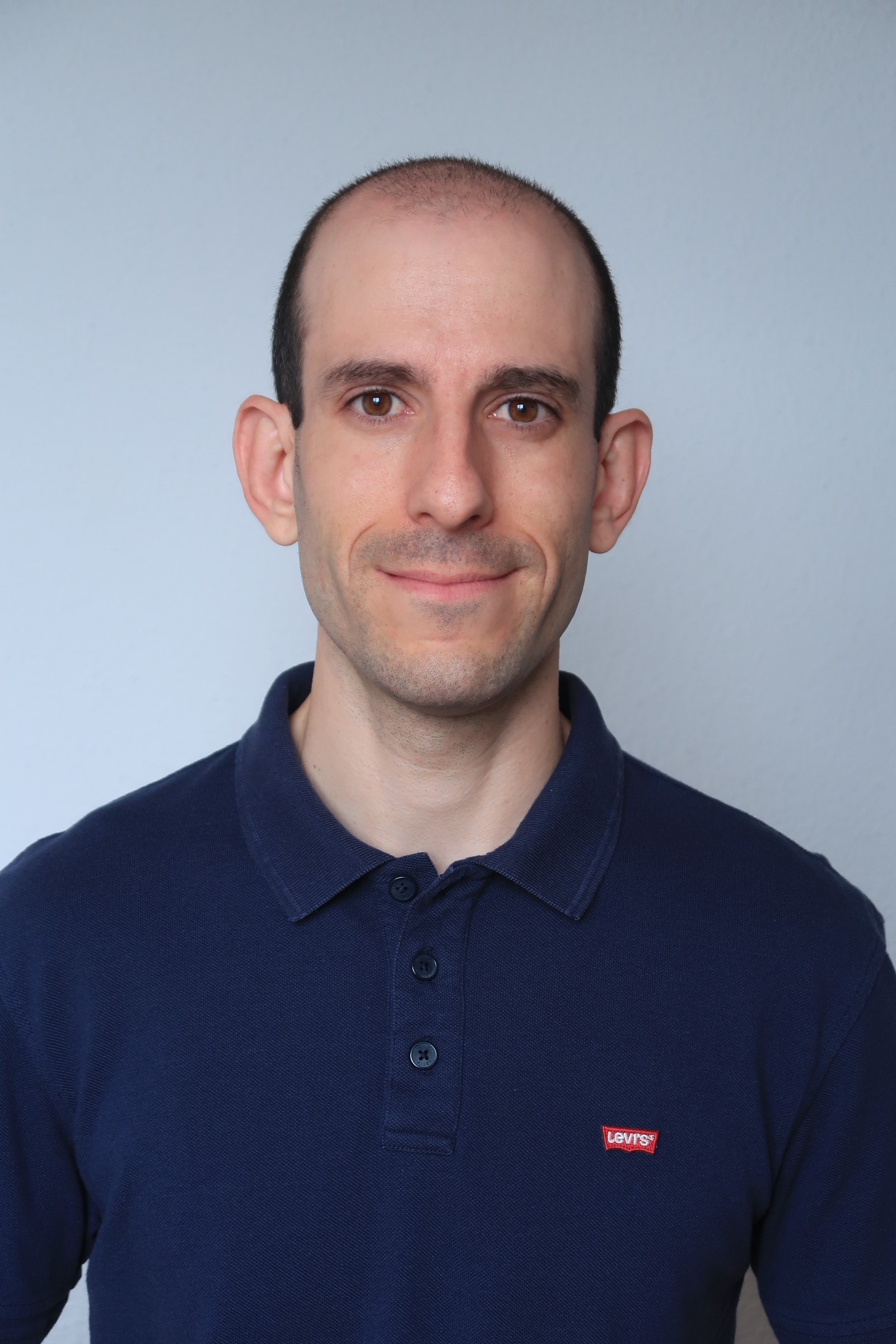}}]
{Ruben Tolosana} received the M.Sc. degree in Telecommunication Engineering, and the Ph.D. degree in Computer and Telecommunication Engineering, from Universidad Autonoma de Madrid, in 2014 and 2019, respectively. In 2014, he joined the Biometrics and Data Pattern Analytics – BiDA Lab at the Universidad Autonoma de Madrid, where he is currently Assistant Professor. He is a member of the ELLIS Society, Technical Area Committee of EURASIP, and Editorial Board of the IEEE Biometrics Council Newsletter. His research interests are mainly focused on signal and image processing, pattern recognition, and machine learning, particularly in the areas of DeepFakes, Human-Computer Interaction, Biometrics, and Health. He is author of more than 80 scientific articles published in international journals and conferences. He has served as General Chair and Program Chair (AVSS 2022), and Area Chair (IJCB 2023, ICPR 2022) in top conferences. Dr. Tolosana has also received several awards such as the European Biometrics Industry Award (2018) from the European Association for Biometrics (EAB) and the Best Ph.D. Thesis Award in 2019-2022 from the Spanish Association for Pattern Recognition and Image Analysis (AERFAI).
\end{IEEEbiography}

\begin{IEEEbiography}[{\includegraphics[width=1in,height=1.25in,clip,keepaspectratio]{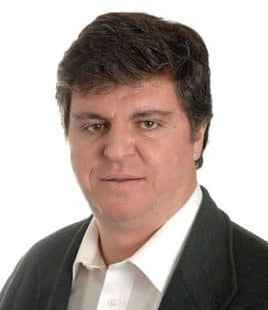}}]
{Javier Ortega-Garcia} received the M.Sc. degree in electrical engineering and the Ph.D. degree (cum laude) in electrical engineering from Universidad Politecnica de Madrid, Spain, in 1989 and 1996, respectively. He is currently a Full Professor at the Signal Processing Chair in Universidad Autonoma de Madrid - Spain, where he holds courses on biometric recognition and digital signal processing. He is a founder and Director of the BiDA-Lab, Biometrics and Data Pattern Analytics Group. He has authored over 300 international contributions, including book chapters, refereed journal, and conference papers. His research interests are focused on biometric pattern recognition (on-line signature verification, speaker recognition, human-device interaction) for security, e-health and user profiling applications. He chaired Odyssey-04, The Speaker Recognition Workshop, ICB-2013, the 6th IAPR International Conference on Biometrics, and ICCST2017, the 51st IEEE International Carnahan Conference on Security Technology.
\end{IEEEbiography}

\EOD

\end{document}